%% file: 00main.tex
\documentclass[12pt]{article}

\usepackage[utf8x]{inputenc}
\usepackage{graphicx}
\usepackage{siunitx}
\DeclareSIUnit\cps{cps}
\usepackage{upgreek}
\usepackage{amsmath}
\usepackage{array}
\usepackage{amssymb}
\usepackage{booktabs}
\usepackage[en-GB]{datetime2}
\usepackage[super]{nth}
\usepackage[labelfont=bf,textfont=md]{caption}
\usepackage{multirow}
\usepackage{lscape}

\usepackage[dvipsnames]{xcolor}

\usepackage{xspace}
\NewDocumentCommand{\nuc}{m o}{\ensuremath{\IfNoValueTF{#2}{}{^{\textrm{#2}}}\textrm{#1}}\xspace}
\NewDocumentCommand{\electron}{}{\ensuremath{e^-}\xspace}
\NewDocumentCommand{\anti}{m}{\ensuremath{\overline{#1}}}
\NewDocumentCommand{\SIadj}{m m}{\SI[number-unit-product={\text{-}}]{#1}{#2}}
\NewDocumentCommand{\be}{}{\ensuremath{\upbeta}\xspace}
\NewDocumentCommand{\mnu}{}{\ensuremath{m_{\upnu}}\xspace}
\NewDocumentCommand{\mnusq}{}{\ensuremath{m_{\upnu}^2}\xspace}

\usepackage{scicite}
\usepackage{times}
\usepackage{setspace}

\makeatletter
\newcommand\footnoteref[1]{\protected@xdef\@thefnmark{\ref{#1}}\@footnotemark}
\makeatother

\newenvironment{nohyphens}
  {\par\hyphenpenalty=10000\exhyphenpenalty=10000\relax}
  {\par}

\newenvironment{sciabstract}{%
\begin{quote} \bf}
{\end{quote}}

\title{Direct neutrino-mass measurement based on 259 days of KATRIN data}
\date{}

\begin{document} 

    \maketitle 
    \input{authors_new}
    
    \input{01abstract}
    \input{02introduction}
    \input{03setup}
    \input{04measurement}
    \input{05analysis}

    \input{06results}
    \input{07discussion}

\begingroup
    \singlespacing
    \bibliography{scibib}
    \bibliographystyle{Science}
\endgroup

    \input{08acknowledgment}

    % just comenting out the part above or below is enough to generate the two documents for the submission
    \newpage
    \input{10supplement}
    
    % \bibliography{scibib}
    % \bibliographystyle{Science}

\end{document}

%% file: authors_new.tex
\renewcommand{\thefootnote}{\fnsymbol{footnote}}

\begin{nohyphens}
\begin{sloppypar}
\noindent
\textbf{KATRIN collaboration}: %\footnote{\todo{add mail address}}:
M.~Aker$^{1}$, 
D.~Batzler$^{1}$, 
A.~Beglarian$^{2}$,  
J.~Behrens$^{1}$, 
J.~Beisenk\"{o}tter$^{3}$, 
M.~Biassoni$^{4}$, 
B.~Bieringer$^{3}$, 
Y.~Biondi$^{1}$, 
F.~Block$^{1}$, 
S.~Bobien$^{5}$, 
M.~B\"{o}ttcher$^{3}$, 
B.~Bornschein$^{1}$, 
L.~Bornschein$^{1}$, 
T.~S.~Caldwell$^{6,7}$, 
M.~Carminati$^{8,9}$, 
A.~Chatrabhuti$^{10}$, 
S.~Chilingaryan$^{2}$, 
B.~A.~Daniel$^{11}$, 
K.~Debowski$^{12}$, 
M.~Descher$^{13}$, 
D.~D\'{i}az~Barrero$^{1}$, 
P.~J.~Doe$^{14}$, 
O.~Dragoun$^{15}$, 
G.~Drexlin$^{13}$, 
F.~Edzards$^{16,17}$, 
K.~Eitel$^{1}$, 
E.~Ellinger$^{12}$, 
R.~Engel$^{1,13}$, 
S.~Enomoto$^{14}$, 
A.~Felden$^{1}$, 
C.~Fengler$^{1}$, 
C.~Fiorini$^{8,9}$, 
J.~A.~Formaggio$^{18}$, 
C.~Forstner$^{16,17}$, 
F.~M.~Fr\"{a}nkle$^{1}$, 
K.~Gauda$^{3}$, 
A.~S.~Gavin$^{6,7}$, 
W.~Gil$^{1}$, 
F.~Gl\"{u}ck$^{1}$, 
S.~Grohmann$^{19}$, 
R.~Gr\"{o}ssle$^{1}$, 
R.~Gumbsheimer$^{1}$, 
N.~Gutknecht$^{13}$, 
V.~Hannen$^{3}$, 
L.~Hasselmann$^{1}$, 
N.~Hau{\ss}mann$^{12}$, 
K.~Helbing$^{12}$, 
H.~Henke$^{1}$, 
S.~Heyns$^{1}$, 
S.~Hickford$^{1}$, 
R.~Hiller$^{1}$, 
D.~Hillesheimer$^{1}$, 
D.~Hinz$^{1}$, 
T.~H\"{o}hn$^{1}$, 
A.~Huber$^{1}$, 
A.~Jansen$^{1}$, 
C.~Karl$^{16}$, 
J.~Kellerer$^{1}$, 
K.~Khosonthongkee$^{20}$, 
M.~Kleifges$^{2}$, 
M.~Klein$^{1}$, 
J.~Kohpei\ss$^{1}$, 
C.~K\"{o}hler$^{16,17}$, 
L.~K\"{o}llenberger$^{1}$, 
A.~Kopmann$^{2}$, 
N.~Kova\v{c}$^{1}$, 
A.~Koval\'{i}k$^{15}$, 
H.~Krause$^{1}$, 
L.~La~Cascio$^{13}$, 
T.~Lasserre$^{21}$, 
J.~Lauer$^{1}$, 
T.~Le$^{1}$, 
O.~Lebeda$^{15}$, 
B.~Lehnert$^{22}$, 
G.~Li$^{11}$, 
A.~Lokhov$^{13,}$\footnote{\label{corresp}Corresponding authors: lokhov@kit.edu, christoph.wiesinger@tum.de}, 
M.~Machatschek$^{1}$, 
M.~Mark$^{1}$, 
A.~Marsteller$^{1,14}$, 
E.~L.~Martin$^{6,7,23}$, 
C.~Melzer$^{1}$, 
S.~Mertens$^{16,17}$, 
S.~Mohanty$^{1}$, 
J.~Mostafa$^{2}$, 
K.~M\"{u}ller$^{1}$, 
A.~Nava$^{4,24}$, 
H.~Neumann$^{5}$, 
S.~Niemes$^{1}$, 
A.~Onillon$^{16,17}$, 
D.~S.~Parno$^{11}$, 
M.~Pavan$^{4,24}$, 
U.~Pinsook$^{10}$, 
A.~W.~P.~Poon$^{22}$, 
J.~M.~Lopez~Poyato$^{25}$, 
S.~Pozzi$^{4}$, 
F.~Priester$^{1}$, 
J.~R\'{a}li\v{s}$,^{15}$, 
S.~Ramachandran$^{12}$, 
R.~G.~H.~Robertson$^{14}$, 
C.~Rodenbeck$^{1,3}$, 
M.~R\"{o}llig$^{1}$, 
C.~R\"{o}ttele$^{1}$, 
M.~Ry\v{s}av\'{y}$,^{15}$, 
R.~Sack$^{1}$, 
A.~Saenz$^{26}$, 
R.~Salomon$^{3}$, 
P.~Sch\"{a}fer$^{1}$, 
M.~Schl\"{o}sser$^{1}$, 
K.~Schl\"{o}sser$^{1}$, 
L.~Schl\"{u}ter$^{17,22}$, 
S.~Schneidewind$^{3}$, 
U.~Schnurr$^{1}$, 
M.~Schrank$^{1}$, 
J.~Sch\"{u}rmann$^{3,26}$, 
A.~Sch\"{u}tz$^{22}$, 
A.~Schwemmer$^{16,17}$, 
A.~Schwenck$^{1}$, 
M.~\v{S}ef\v{c}\'{i}k$^{15}$, 
D.~Siegmann$^{16,17}$, 
F.~Simon$^{2}$, 
F.~Spanier$^{27}$, 
D.~Spreng$^{16,17}$, 
W.~Sreethawong$^{20}$, 
M.~Steidl$^{1}$, 
J.~\v{S}torek$^{1}$, 
X.~Stribl$^{16,17}$, 
M.~Sturm$^{1}$, 
N.~Suwonjandee$^{10}$, 
N.~Tan~Jerome$^{2}$, 
H.~H.~Telle$^{25}$, 
L.~A.~Thorne$^{28}$, 
T.~Th\"{u}mmler$^{1}$, 
S.~Tirolf$^{1}$, 
N.~Titov$^{29}$, 
I.~Tkachev$^{29}$, 
K.~Urban$^{16,17}$, 
K.~Valerius$^{1}$, 
D.~V\'{e}nos$^{15}$, 
C.~Weinheimer$^{3}$, 
S.~Welte$^{1}$, 
J.~Wendel$^{1}$, 
C.~Wiesinger$^{16,17,}$\footnoteref{corresp}, %\footnote{Corresponding author: christoph.wiesinger@tum.de}, 
J.~F.~Wilkerson$^{6,7,}$\footnote{Also affiliated with Oak Ridge National Laboratory, Oak Ridge, TN 37831, USA}, 
J.~Wolf$^{13}$, 
S.~W\"{u}stling$^{2}$, 
J.~Wydra$^{1}$, 
W.~Xu$^{18}$, 
S.~Zadorozhny$^{29}$, 
G.~Zeller$^{1}$
\end{sloppypar}
\end{nohyphens}

\vspace{1em}

\begin{sloppypar}
\begin{flushleft}
\noindent
%\scriptsize{
$^{1}$ Institute for Astroparticle Physics~(IAP), Karlsruhe Institute of Technology~(KIT), Hermann-von-Helmholtz-Platz 1, 76344 Eggenstein-Leopoldshafen, Germany \\
$^{2}$ Institute for Data Processing and Electronics~(IPE), Karlsruhe Institute of Technology~(KIT), Hermann-von-Helmholtz-Platz 1, 76344 Eggenstein-Leopoldshafen, Germany \\
$^{3}$ Institute for Nuclear Physics, University of M\"{u}nster, Wilhelm-Klemm-Str.~9, 48149 M\"{u}nster, Germany \\
$^{4}$ Istituto Nazionale di Fisica Nucleare (INFN) -- Sezione di Milano-Bicocca, Piazza della Scienza 3, 20126 Milano, Italy \\
$^{5}$ Institute for Technical Physics~(ITEP), Karlsruhe Institute of Technology~(KIT), Hermann-von-Helmholtz-Platz 1, 76344 Eggenstein-Leopoldshafen, Germany \\
$^{6}$ Department of Physics and Astronomy, University of North Carolina, Chapel Hill, NC 27599, USA \\
$^{7}$ Triangle Universities Nuclear Laboratory, Durham, NC 27708, USA \\
$^{8}$ Politecnico di Milano, Dipartimento di Elettronica, Informazione e Bioingegneria, Piazza L. da Vinci 32, 20133 Milano, Italy \\
$^{9}$ Istituto Nazionale di Fisica Nucleare (INFN) -- Sezione di Milano, Via Celoria 16, 20133 Milano, Italy \\
$^{10}$ Department of Physics, Faculty of Science, Chulalongkorn University, Bangkok 10330, Thailand \\
$^{11}$ Department of Physics, Carnegie Mellon University, Pittsburgh, PA 15213, USA \\
$^{12}$ Department of Physics, Faculty of Mathematics and Natural Sciences, University of Wuppertal, Gau{\ss}str.~20, 42119 Wuppertal, Germany \\
$^{13}$ Institute of Experimental Particle Physics~(ETP), Karlsruhe Institute of Technology~(KIT), Wolfgang-Gaede-Str.~1, 76131 Karlsruhe, Germany \\
$^{14}$ Center for Experimental Nuclear Physics and Astrophysics, and Dept.~of Physics, University of Washington, Seattle, WA 98195, USA \\
$^{15}$ Nuclear Physics Institute,  Czech Academy of Sciences, 25068 \v{R}e\v{z}, Czech Republic \\
$^{16}$ Technical University of Munich, TUM School of Natural Sciences, Physics Department, James-Franck-Stra\ss e 1, 85748 Garching, Germany \\
$^{17}$ Max Planck Institute for Physics, Boltzmannstr. 8, 85748 Garching, Germany \\
$^{18}$ Laboratory for Nuclear Science, Massachusetts Institute of Technology, 77 Massachusetts Ave, Cambridge, MA 02139, USA \\
$^{19}$ Karlsruhe Institute of Technology (KIT), Institute for Technical Thermodynamics and Refrigeration, Engler-Bunte-Ring 21, 76131 Karlsruhe, Germany \\
$^{20}$ School of Physics and Center of Excellence in High Energy Physics and Astrophysics, Suranaree University of Technology, Nakhon Ratchasima 30000, Thailand \\
$^{21}$ IRFU (DPhP \& APC), CEA, Universit\'{e} Paris-Saclay, 91191 Gif-sur-Yvette, France \\
$^{22}$ Nuclear Science Division, Lawrence Berkeley National Laboratory, Berkeley, CA 94720, USA \\
$^{23}$ Current address: Department of Physics, Duke University, Durham, NC, 27708, USA  \\
$^{24}$ Dipartimento di Fisica, Universit\`{a} di Milano - Bicocca, Piazza della Scienza 3, 20126 Milano, Italy \\
$^{25}$ Departamento de Qu\'{i}mica F\'{i}sica Aplicada, Universidad Autonoma de Madrid, Campus de Cantoblanco, 28049 Madrid, Spain \\
$^{26}$ Institut f\"{u}r Physik, Humboldt-Universit\"{a}t zu Berlin, Newtonstr.~~15, 12489 Berlin, Germany \\
$^{27}$ Institute for Theoretical Astrophysics, University of Heidelberg, Albert-Ueberle-Str.~2, 69120 Heidelberg, Germany \\
$^{28}$ Institut f\"{u}r Physik, Johannes-Gutenberg-Universit\"{a}t Mainz, 55099 Mainz, Germany \\
$^{29}$ Institute for Nuclear Research of Russian Academy of Sciences, 60th October Anniversary Prospect 7a, 117312 Moscow, Russia\footnote{Institutional status in the KATRIN collaboration has been suspended since February 24, 2022} \\
%$^{30}$ Also affiliated with Oak Ridge National Laboratory, Oak Ridge, TN 37831, USA \\
%}
\end{flushleft}
\end{sloppypar}

\renewcommand{\thefootnote}{\arabic{footnote}}
\setcounter{footnote}{0}

%% file: 01abstract.tex
\begin{sciabstract} 
    The fact that neutrinos carry a non-vanishing rest mass is evidence of physics beyond the Standard Model of elementary particles. Their absolute mass bears important relevance from particle physics to cosmology. In this work, we report on the search for the effective electron antineutrino mass with the KATRIN experiment. KATRIN performs precision spectroscopy of the tritium \be-decay close to the kinematic endpoint. Based on the first five neutrino-mass measurement campaigns, we derive a best-fit value of \mbox{$\mnu^{2} = \SI[parse-numbers = false]{-0.14^{+0.13}_{-0.15}}{\electronvolt\squared}$}, resulting in an upper limit of \mbox{$\mnu < \SI{0.45}{\electronvolt}$} at \SI{90}{\percent} confidence level. With six times the statistics of previous data sets, amounting to \num{36} million electrons collected in \num{259} measurement days, a substantial reduction of the background level and improved systematic uncertainties, this result tightens KATRIN's previous bound by a factor of almost two.
\end{sciabstract}

%% file: 02introduction.tex
\section*{Introduction}
\label{Sec:Introduction}

The discovery of neutrino flavor oscillations implies the existence of different mass states and contradicts the hypothesis of massless neutrinos in the Standard Model of elementary particles~\cite{Super-Kamiokande:1998kpq, SNO:2002tuh}. While the squared mass splittings are measured with ever-increasing precision~\cite{Esteban:2020cvm, Capozzi:2021fjo}, the absolute neutrino mass scale remains unknown. Neutrinos are at least six orders of magnitude lighter than other fermions, suggesting a new type of mass-generation mechanism that could involve heavy non-active neutrinos~\cite{Minkowski:1977sc, Gell-Mann:1979vob, Mohapatra:1979ia, Yanagida:1979as}.

Neutrinos come in three flavors, $\upnu_l$, named according to their corresponding charged-lepton partners, $l \in \{e, \upmu, \uptau\}$~\cite{Danby:1962nd, DONUT:2000fbd}. Neutrino flavor oscillations identify them as superpositions of individual mass states, $\upnu_i$, $i \in \{1, 2, 3\}$, with mass values $m_i$. The Pontecorvo-Maki-Nakagawa-Sakata (PMNS) mixing matrix elements $U_{li}$ encode the $\upnu_i$-admixture of $\upnu_l$ \cite{Maki:1962mu,Pontecorvo:1967fh}.

Neutrinos are the most abundant massive known particles in our universe. Their late transition from hot to cold matter leaves an imprint on the cosmic evolution and structure formation. The combination of several observational data sets, analyzed within the framework of standard cosmology ($\Lambda$CDM), provides a stringent bound of \SI{0.072}{\electronvolt} at \SI{95}{\percent} credible interval (CI)~\cite{Planck:2018vyg,DESI:2024mwx} for the neutrino mass sum, $\sum_i m_i$.%
\footnote{In this paper we use natural units, meaning $c=1$.}
This bound can be relaxed in the case of non-standard physics, e.g.~\cite{Farzan:2015pca,Oldengott:2019lke,Escudero:2020ped}.

Neutrinos carry neither electric nor color charges. Hence, they may be Majorana particles, i.e.\ fermions that are their own antiparticles~\cite{Majorana:1937vz}. 
This hypothesis is tested by experiments searching for neutrinoless double-\be decay~\cite{Schechter:1981bd}. Assuming mediation by light Majorana neutrinos, upper limits on the effective Majorana neutrino mass, $| \sum_i U_{ei}^2 m_i |$, the coherent sum of neutrino masses weighted by their squared electron neutrino contribution, are placed at \SI[parse-numbers = false]{0.079-0.180}{\electronvolt} at \SI{90}{\percent} confidence level (CL) for \nuc{Ge}[76]~\cite{GERDA:2020xhi}, \SI[parse-numbers = false]{0.070-0.240}{\electronvolt} at \SI{90}{\percent} CI for \nuc{Te}[130]~\cite{CUORE:2024ikf}, and \SI[parse-numbers = false]{0.036-0.156}{\electronvolt} at \SI{90}{\percent} CL for \nuc{Xe}[136]~\cite{KamLAND-Zen:2022tow}.\footnote{The ranges correspond to different nuclear structure calculations.}

A direct way to assess the neutrino mass is provided by \be-decay kinematics, where the creation of (anti-)neutrinos modifies the phase space available for electron emission.\footnote{Similar considerations apply to electron-capture processes~\cite{Gastaldo:2022uyk}.} This shape distortion is maximal near the kinematic endpoint of the \be-decay energy spectrum. As each mass state contributes individually with its own admixture but is typically not resolved, the squared effective electron antineutrino mass, $\mnu^2 = \sum_i |U_{ei}|^2 m_i^2$, is probed.%
\footnote{In this paper we use the term neutrino mass for $\mnu$. Some sources use the symbol $m_\upbeta$.}
Unlike the above-mentioned methods, this approach is independent of both the cosmological model and the neutrino nature.

The KArlsruhe TRItium Neutrino experiment (KATRIN) leads the direct neutrino mass exploration. Based on the first measurement campaign, long-standing bounds on the neutrino mass obtained by the previous Mainz and Troitsk experiments~\cite{Kraus:2005,Aseev:2011} were improved by a factor of two~\cite{KATRIN:2019yun}. The addition of the second campaign resulted in the world's first sub-\si{\electronvolt} constraint of $\mnu < \SI{0.8}{\electronvolt}$ at \SI{90}{\percent} CL~\cite{KATRIN:2021uub}, which was based on six million electron events in the analysis region. In this work, we present the result of the first five measurement campaigns with \num{36} million electrons collected over \num{259} measurement days.

%% file: 03setup.tex
\section*{Experimental setup} 
\label{Sec:ExperimentalSetup}

\begin{figure}[!t]
    \centering
    \includegraphics[width=1.0\textwidth]{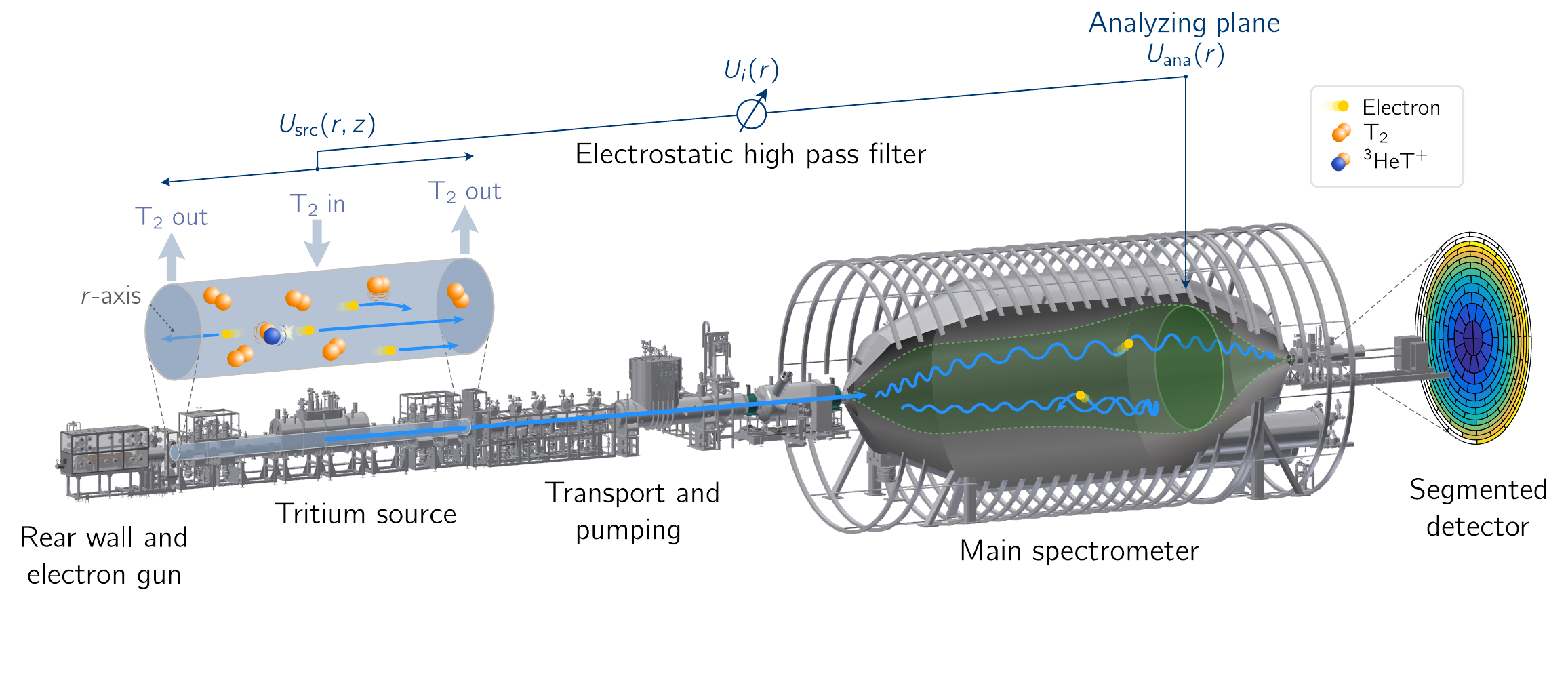}
    \caption{The KATRIN beamline. Tritium gas is continuously injected into the source, where it decays, producing \be-electrons. The inset depicts the source volume, filled with a plasma of low-energy electrons and tritiated ions. The tritium gas is pumped out, purified, and re-circulated, while the electrons are guided magnetically through the transport section into the spectrometer. Only electrons with sufficient energy to overcome the electrostatic potential in the analyzing plane are counted at the detector. The analyzing plane is shifted towards the detector for background reduction (see main text for explanation).}
    \label{Fig:Beamline}
\end{figure}

KATRIN performs precision \be-spectroscopy close to the kinematic endpoint, $E_0 \approx$ \SI{18.6}{\kilo\electronvolt}, of molecular tritium decay
\begin{equation}
    \nuc{T}_2 \rightarrow\ \nuc{He}[3]\nuc{T}^+ + \electron + \anti{\upnu}_e.
\end{equation}
It combines a high-activity gaseous tritium source of up to \SI{100}{\giga\becquerel} with a high-resolution spectrometer~\cite{KATRIN:2021dfa}. The \SIadj{70}{\meter}-long KATRIN beamline is illustrated in figure~\ref{Fig:Beamline}.

Tritium is continuously reprocessed by the Tritium Laboratory Karlsruhe (TLK), delivering a tritium throughput of up to \SI{40}{\gram/\day} with up to \SI{99}{\percent} isotopic purity to the KATRIN source~\cite{Hillesheimer2024_Tritium}. Starting at a magnetic field of about \SI{2.5}{\tesla}, the \be-decay electrons are guided adiabatically through the windowless source and the transport section, where the tritium is removed by differential and cryogenic pumping~\cite{Marsteller:2020tgj,Rottele:2023cgk}. The upstream flux of electrons is terminated with a gold-plated rear wall. A voltage of $\mathcal{O}(\SI{100}{\milli\volt})$ is applied to the rear wall to control the source potential~\cite{KATRIN:2021dfa}.

The spectrometer section consists of a pre-spectrometer followed by the \SIadj{23}{\meter}-long, \SIadj{10}{\meter}-wide main spectrometer, both using the magnetic adiabatic collimation with electrostatic filtering (MAC-E) principle~\cite{Lobashev:1985mu,Picard_1992345}. 
The electron momenta, $\vec{p}=\vec{p}_\perp+\vec{p}_\parallel$, are aligned with the field lines of a magnetic field relaxing in strength, $B$, due to the conservation of the electron orbital magnetic moment, $\mu\propto \frac{p^2_{\perp}}{B}$. In the main spectrometer, the magnetic field strength decreases to about \SI{0.6}{\milli\tesla} at its minimum. The simultaneous application of a precisely known electrostatic retarding potential~\cite{Rodenbeck:2022iys}, $U$, allows only electrons of charge $q=-e$ and energy $E > qU$ to pass the so-called analyzing plane, where maximal collimation and retardation coincide.
With the maximal magnetic field in the beamline at about \SI{4.2}{\tesla}, an acceptance of electrons emitted with a pitch angle up to \SI{51}{\degree} and an excellent energy filter width of $\mathcal{O}(\SI{1}{\electronvolt})$ are achieved. The integral flux of transmitted electrons is measured by the focal plane detector (FPD), a 148-pixel silicon-PIN-diode array, featuring a detection efficiency of about \SI{95}{\percent} \cite{Amsbaugh:2014uca}.

KATRIN is monitored by more than \num{5700} sensors, recording numerous operational parameters such as the tritium purity, the temperatures of different beamline components, and the magnetic fields~\cite{KATRIN:2021dfa}. Additionally, the beamline is equipped with several calibration sources. An angular-selective source of monoenergetic photoelectrons (electron gun)~\cite{Behrens:2017cmd} is used to measure the gas density in the tritium source down to sub-percent precision. The co-circulation of \nuc{Kr}[83m] with tritium and other carrier gases~\cite{Sentkerestiova_2018,KATRIN:2022zqa} allows us to determine the electron starting potential within the source and the electromagnetic fields in the spectrometer.

Despite the high-activity tritium source, the signal count rate is less than one count per second (\si{\cps}) at $qU \gtrsim E_0 - \SI{10}{\electronvolt}$, which puts stringent requirements on the background rate. External backgrounds due to gamma radiation and cosmic-ray muons are mitigated by the magnetic shielding and the electric potential on the wire electrode system of the MAC-E spectrometers~\cite{KATRIN:2021dfa,KATRIN:2018rxw,KATRIN:2019dnj}. The \nuc{Rn}[219] and \nuc{Rn}[220] decay backgrounds are successfully suppressed by liquid nitrogen-cooled copper baffles in the main-spectrometer pumping ducts~\cite{Wandkowsky:2013una,Gorhardt:2018rqg}. Particles stored in a Penning trap between the two spectrometers generate another source of background. These accumulated particles are removed by a conductive wire periodically swiping the trap volume~\cite{KATRIN:2019mkh}. The background contribution of the detector system is suppressed by a region-of-interest (ROI) cut on the detected energy, a muon veto system, and post-acceleration of the electrons~\cite{Amsbaugh:2014uca}.

The remaining background is dominated by neutral atoms in highly excited states, which are sputtered off the spectrometer's inner surface by the decays of residual \nuc{Pb}[210] and its daughter \nuc{Po}[210]. These atoms are distributed throughout the spectrometer volume. Their loosely bound electrons are easily released by blackbody radiation or by autoionization and are accelerated up to signal-electron energies towards the detector~\cite{KATRIN:2020zld}. A method to reduce this background is a re-configuration of the electromagnetic fields in the main spectrometer. It comprises both a compression of the magnetic flux tube and a downstream shift of the analyzing plane, which is nominally situated in the center of the main spectrometer~\cite{Lokhov:2022iag}. This shifted-analyzing-plane (SAP) setting was implemented over the course of the new measurement campaigns, see Tab.~\ref{Tab:MeasurementCampaigns}.

%% file: 04measurement.tex
\section*{Measurement overview} 
\label{Sec:MeasurementOverview}

KATRIN obtains the integral \be-spectrum by measuring the count rate at the FPD for a defined sequence of retarding-energy set points, $qU_i$, forming a \be-spectrum scan. Each scan is typically composed of up to \num{40} set points in a range of $E_0 - \SI{300}{\electronvolt} \leq qU_i \leq E_0 + \SI{135}{\electronvolt}$. The neutrino mass analysis is limited to data points above $E_0 -$ \SI{40}{\electronvolt}. The spectrum points above the endpoint constrain the background rate. The data recorded below the endpoint, outside of the analysis range, is used for calibration and monitoring purposes.
The total duration of a \be-spectrum scan is about three hours, while the fraction of time spent at each scan step is optimized for neutrino mass sensitivity. The sequence of retarding-energy set points is increasing, decreasing, or randomly set.  

Each KATRIN Neutrino Mass (KNM) measurement campaign contains a few hundred \be-spectrum scans. In this work, we present the first five campaigns, with the data taken between March 2019 and June 2021. Table \ref{Tab:MeasurementCampaigns} contains the relevant operational information. The key configuration changes are:
\begin{itemize}
    \item After an initial burn-in period, the tritium gas column density was raised from $\rho d =$ \SI{1.08e21}{\per\metre\squared} in KNM1 to \SI{4.20e21}{\per\metre\squared} in KNM2~\cite{KATRIN:2021uub},  corresponding to \SI{84}{\percent} of the design value. $\rho$ is the average gas density and $d =$ \SI{10}{\meter} is the length of the source. 
    \item Starting from KNM3 the background was reduced by a factor of two, using the shifted-analyzing-plane setting~\cite{Lokhov:2022iag}. We validated the feasibility of this configuration by \be-spectrum scans in the shifted (KNM3-SAP) and nominal symmetrical (KNM3-NAP) analyzing-plane settings. 
    \item A powerful technique to characterize the electric-potential variations in the source with \nuc{Kr}[83m] conversion electrons was introduced after KNM3~\cite{KATRIN:2022zqa}. This method features a new co-circulation mode of krypton with tritium at a high column density of $\rho d = $ \SI{3.8e21}{\per\metre\squared}. This column density was chosen as the nominal setting after KNM3-NAP in order to perform the \be-spectrum scans under the same experimental conditions. The source temperature was raised from \SI{30}{\kelvin} in KNM2 to \SI{80}{\kelvin} in KNM3-SAP to allow for gaseous krypton co-circulation.
    \item The impact of the time-dependent accumulation of particles in the Penning trap between the two spectrometers was 
    reduced during KNM4 by shorter intervals between swipings of the trap and was fully mitigated by reducing the pre-spectrometer potential. In addition, the time spent at each scan step was further optimized for a better neutrino mass sensitivity from the nominal (KNM4-NOM) to the optimized (KNM4-OPT) configuration.
    \item Over the course of the first four campaigns the gold surface of the rear wall accumulated tritium that produces additional \be-decay electrons. The corresponding activity is proportional to an integral tritium flow of \SI{2.9e7}{\milli\bar\cdot l}. Prior to the start of KNM5, the accumulated tritium was reduced by a factor of \num{e3} by ozone cleaning of the rear wall~\cite{Aker2023:OzonCleaning}.
\end{itemize}

\begin{landscape}
\vspace*{\fill}
\begin{table}[!ht]
    \centering
    \caption{Key features of the measurement campaigns. The live time corresponds to the data used for analysis. The values are quoted for the number of active pixels in each measurement campaign and all the scan-steps above $E_0 -$ \SI{40}{\electronvolt}.}
    \begin{tabular}{lccccccc}
        \toprule
        & \textbf{KNM1} & \textbf{KNM2} & \textbf{KNM3-SAP} & \textbf{KNM3-NAP} & \textbf{KNM4-NOM} & \textbf{KNM4-OPT} & \textbf{KNM5} \\
        \midrule
        Year                          
                & 2019 & 2019 & 2020 & 2020 & 2020 & 2020 & 2021 \\
        Month                          
                & 04 -- 05 & 10 -- 12 & 06 -- 07 & 07 & 09 -- 10 & 10 -- 12 & 04 -- 06 \\
        Measurement days                         
                & \num{35} & \num{45} & \num{14} & \num{14} & \num{49} & \num{30} & \num{72} \\
        \midrule
        Analyzing plane
                & nominal & nominal & shifted & nominal & shifted & shifted & shifted \\
        Pre-spectrometer voltage (\si{\kilo\volt})
                & \num{-10.5} & \num{-10.5} & \num{-10.5} & \num{-10.5} & \num{-10.5} & \num{-10.5}/\num{-0.1} & \num{-0.1} \\
        Scan-step duration 
                & nominal & nominal & nominal & nominal & nominal & optimized & optimized \\
        Rear-wall
                & - & - & - & - & - & - & cleaned \\
        Source temperature (\si{\kelvin})
                & \num{30} &  \num{30} &  \num{80} &  \num{80} &  \num{80} &  \num{80} &  \num{80} \\
        Background rate (\si{cps})              
                & \num{0.29} & \num{0.22} & \num{0.12} & \num{0.22} & \num{0.13} & \num{0.13} & \num{0.14} \\
        \midrule
        Number of scans                          
                & \num{274} & \num{361} & \num{114} & \num{116} & \num{320} & \num{150} & \num{422} \\
        Live time (\si{hrs})                     
                & \num{522} & \num{694} & \num{220} & \num{224} & \num{835} & \num{432} & \num{1226} \\
        Active pixels                     
                & \num{117} & \num{117} & \num{126} & \num{126} & \num{126} & \num{126} & \num{126} \\
        Column density ($\times$\SI{e21}{\per\metre\squared})
                & \num{1.08} & \num{4.20} & \num{2.05} & \num{3.70} & \num{3.76} & \num{3.76} & \num{3.77} \\
        Source activity (\si{\giga\becquerel})        
                & \num{24} & \num{94} & \num{46} & \num{83} & \num{84} & \num{84} & \num{84}\\
        Number of counts ($\times$\num{e6})
                & \num{2.03} & \num{4.31} & \num{1.07} & \num{1.43} & \num{5.64} & \num{4.58} & \num{16.65} \\
        \bottomrule
    \end{tabular}
    \label{Tab:MeasurementCampaigns}
\end{table}
\vspace*{\fill}
\end{landscape}

%% file: 05analysis.tex
\section*{Data analysis} 
\label{Sec:AnalysisStrategy}

In total, \num{1757} out of \num{1895} scans with \num{117} (in KNM1 and 2), later \num{126} (in KNM3-SAP, 3-NAP, 4-NOM, 4-OPT, and 5), out of \num{148} pixels were selected for the analysis.\footnote{Nine pixels were recovered after the exchange of the detector wafer.} This selection excludes scans in which the monitoring systems indicate instabilities in electromagnetic fields or source parameters, and pixels which were shadowed by structural material of the beamline. The total number of counts in the analysis range is \num{36} million, which includes both signal electrons and background. Compared to the previous result reported in~\cite{KATRIN:2021uub} the statistics improved by a factor of six. 

The high reproducibility of the retarding-energy set points (with a standard deviation of \SI{<10}{\milli\volt}) and the stability of the operational parameters allows the combination of the \be-spectrum scans within one campaign by summing the counts recorded at a given set point. Analogously, the statistically independent spectra recorded by different detector pixels are combined. For campaigns operated in the nominal-analyzing-plane configuration the counts of all pixels are summed. For the shifted-analyzing-plane configuration the pixels are grouped into \num{14} ring-like detector patches, to account for the electric-potential and magnetic-field variations in the analyzing plane. This data combination procedure facilitates the analysis while introducing negligible additional uncertainties.

The expected rate $R_\mathrm{calc}\left(qU_i\right)$ at each retarding-energy set point, $qU_i$, is derived from the theoretical tritium \be-decay energy spectrum, $R_\upbeta(E; E_0, \mnu^2)$, convolved with the experimental response function, $f_\mathrm{calc}(E, qU_i)$,
\begin{equation}
    \label{Ntheo}
    R_{\mathrm{calc}}\left(qU_i\right) = A_{\mathrm{s}} N_{\mathrm{T}} \int_{qU_i}^{\infty}R_\upbeta\left(E; E_0, \mnu^2\right) f_\mathrm{calc}\left(E, qU_i\right) \ \mathrm{d} E + R_{\mathrm{bg}}.
\end{equation} 
The theoretical \be-decay spectrum is calculated with Fermi’s golden rule~\cite{Kleesiek:2018mel}. Molecular effects are encoded in the final-states distribution, which describes the probability to populate different molecular states by the decay of molecular tritium and is obtained by ab initio calculations~\cite{Saenz:2000dul,Doss:2006zv}.
This effect is included in $R_\upbeta$.

The response function describes the transmission probability of electrons through the MAC-E filter and takes into account the energy losses due to inelastic scattering on tritium molecules in the source. $N_{\mathrm{T}}$ depends on the number of tritium atoms in the source, the maximum acceptance angle of the MAC-E filter, and the detection efficiency. The normalization factor $A_{\mathrm{s}}$, the background rate $R_{\mathrm{bg}}$, the effective endpoint energy $E_0$, and the squared neutrino mass $\mnu^2$ are free parameters of the model.  

In order to take into account the different experimental conditions, each campaign is described with an individual model. Likewise, the spectra recorded by different detector patches are described by individual models. The parameter inference is performed by a combined maximum-likelihood fit to statistically independent data sets $p$, minimizing the sum of the negative logarithm of the likelihoods, $-2 \log \mathcal{L} = -2 \sum_{p} \log \mathcal{L}_{p}$. For the campaigns operated in the nominal-analyzing-plane configuration, we use the standard normal likelihood function, assuming a Gaussian distribution of the measured counts.
For campaigns in the shifted analyzing plane configuration, where the subdivision of data into detector patches leads to a \num{14}-times-lower count rate, we use the Poisson likelihood function. 

In order to include systematic uncertainties, we extend the likelihood function with pull terms of the form $\left({\vec{\eta}}-\vec{\eta}_\mathrm{ext}\right)^\mathrm{T} \Theta_\mathrm{cov}^{-1} \left( {\vec{\eta}}-\vec{\eta}_\mathrm{ext}\right)$. This allows the experimental parameters, $\vec{\eta}$, such as the magnetic fields and tritium column density, to vary according to the covariance matrix, $\Theta_\mathrm{cov}$, around our best external estimate, $\vec{\eta}_\mathrm{ext}$, typically obtained from dedicated calibration measurements.

Due to the high degree of data segmentation, resulting in a total of 1609 data points, each minimization step requires $\mathcal{O}(10^3)$ computationally expensive evaluations of $R_{\mathrm{calc}}\left(qU_i\right)$. To overcome this computational challenge we employ two methods: a highly optimized direct calculation of the model~\cite{Kleesiek:2018mel}, and a fast model prediction with a neural network~\cite{Karl:2022jda}.

To eliminate experimenter's bias, the analysis is carried out by two independent teams applying a two-step approach. Firstly, the full analysis is performed on simulated data sets mimicking the recorded experimental conditions of each scan. Secondly, when analyzing the data we use a model blinding scheme in which we alter the variance of the molecular final-states distribution by an unknown value, so that \mnusq acquires an unknown bias. Only after the analysis procedure and input parameters are fixed, the unmodified true final-states distribution is used.

This step was performed twice. 
Post-unblinding, a mistake in the data combination of the KNM4 campaign was uncovered.
The change of the relative scan-step durations, in combination with a drift of the starting potential, requires this data set to be separated into two periods, KNM4-NOM and KNM4-OPT.
Additionally, a thorough review of all analysis inputs was repeated. The following modifications were made:
the model of the time-dependent background was changed to a non-linear description, motivated by simulations and additional measurements;
the uncertainties of the energy loss function were re-evaluated; and the energy-dependent angle of the monoenergetic photoelectron source was included in the column density evaluation.
The change of the best-fit \mnusq value attributed to these modifications is small compared to the uncertainty, while the systematic uncertainties have increased. The details are provided in the supplementary material.

%% file: 06results.tex
\section*{Results} 
\label{Sec:Results}

\begin{figure}[!t]
    \centering
    \includegraphics[width=0.6\textwidth]
    {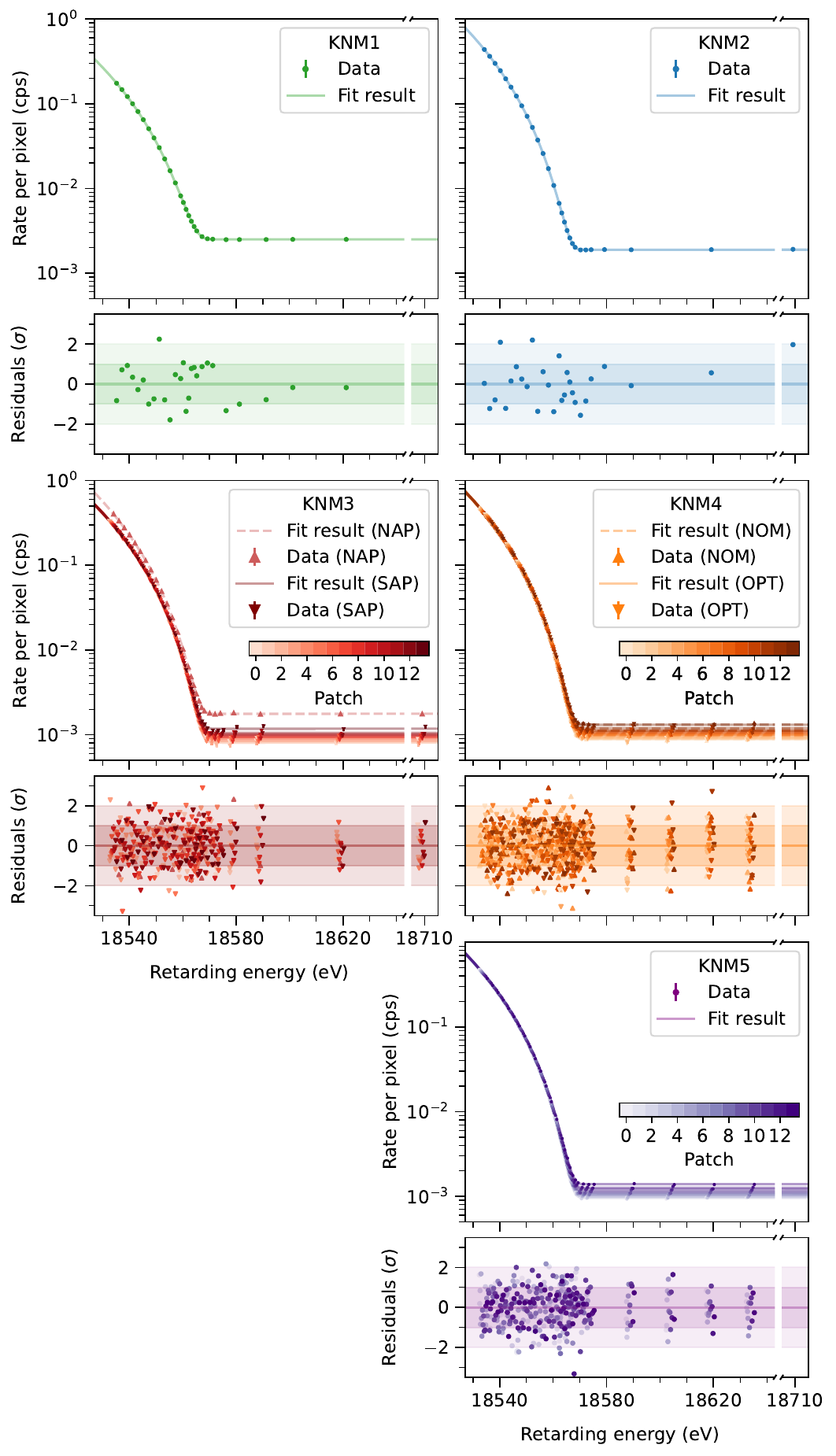}
    \caption{Spectra, fit models and normalized residuals of each measurement campaign. The KNM3-SAP, KNM4-NOM, and KNM4-OPT, and KNM5 data are subdivided into \num{14} detector patches. The squared neutrino mass is a common fit parameter over all data sets.}
    \label{Fig:Spectra}
\end{figure}

The simultaneous fit of the first five measurement campaigns yields a squared neutrino mass $\mnu^{2} = \SI[parse-numbers = false]{-0.14^{+0.13}_{-0.15}}{\electronvolt\squared}$ with an excellent goodness-of-fit. The corresponding p-value is \num{0.84}. Negative $\mnu^{2}$ estimates due to statistical fluctuations are allowed by the spectrum model. Both analysis teams obtain the same result within \SI{4}{\percent} of the total \mnusq uncertainty. The spectra of each campaign and each detector patch are shown in figure~\ref{Fig:Spectra}. The total uncertainty is dominated by the statistical error, followed by the uncertainties of the column density, the energy-loss function, the time-dependent background rate, and the source-potential variations. The uncertainties are listed in table~\ref{Tab:SystematicBreakdown}.

\begin{table}[!b]
    \centering
    \caption{Breakdown of the uncertainties based on the Asimov data set. Each contribution in the last block of the table is smaller than \SI{0.002}{\electronvolt\squared} and they are therefore not propagated into the fit. Detailed information about the various effects is provided in the supplementary material.}
    \begin{tabular}{lc}
        \toprule
        \textbf{Effect} & \textbf{\SI{68.3}{\percent} CL uncertainty on $\boldsymbol{m_{\upnu}^{2}}$ (\si{\electronvolt\squared})} \\
        \midrule
        Statistical uncertainty                               & \num{0.108} \\ 
        Non-Poissonian background                             & \num{0.015} \\ 
        \midrule
        Column density $\times$ inelastic cross section       & \num{0.052} \\ 
        Energy-loss function                                  & \num{0.034} \\ 
        Scan-step-duration-dependent background               & \num{0.027} \\ 
        Source-potential variations                           & \num{0.022} \\
        $qU$-dependent background slope                       & \num{0.007} \\ 
        Analyzing-plane magnetic field and potential          & \num{0.006} \\ 
        Source magnetic field                                 & \num{0.004} \\ 
        Maximum magnetic field                                & \num{0.004} \\
        Rear-wall residual tritium background                 & \num{0.004} \\
        \midrule
        Molecular final-state distribution                    & \multirow{5}{*}{\num{<0.002}} \\
        Activity fluctuations                                 & \\
        Detector efficiency                                   & \\
        Retarding-potential stability and reproducibility     & \\
        Theoretical corrections                               & \\
        \bottomrule
    \end{tabular}
    \label{Tab:SystematicBreakdown}
\end{table}

Based on this best-fit result we obtain an upper limit of $\mnu < \SI{0.45}{\electronvolt}$ at \SI{90}{\percent} CL, using the Lokhov-Tkachov method~\cite{LokhovTkachov:2015}. By construction, this upper limit equals the sensitivity of the experiment. This technique avoids shrinking upper limits for more negative values of $\mnu^{2}$, which is characteristic for other methods, such as the Feldman-Cousins method~\cite{FeldmanCousins:1998}. For completeness, the upper limit from the latter is $\mnu <$ \SI{0.31}{\electronvolt} at \SI{90}{\percent} CL.

The effective endpoint, $E_{0}$, is directly related to the Q-value of the tritium \be-decay when taking into account the work function differences between the source, the rear wall and the spectrometer, as well as the recoil energy of the molecular ion. The Q-value for T$_2$ $\upbeta$-decay measured in the campaign with the best source-potential stability, KNM4-NOM, is \SI{18575.0\pm0.3}{\electronvolt}. This agrees with our previously published results~\cite{KATRIN:2019yun,KATRIN:2021dfa}, but is in slight tension with independent measurements of the atomic-mass difference with precision Penning traps~\cite{MedinaRestrepo:2023qbj}.

%% file: 07discussion.tex
\section*{Discussion}
\label{Sec:Conclusion}

\begin{figure}[!ht]
    \centering
    \includegraphics[width=0.6\textwidth]{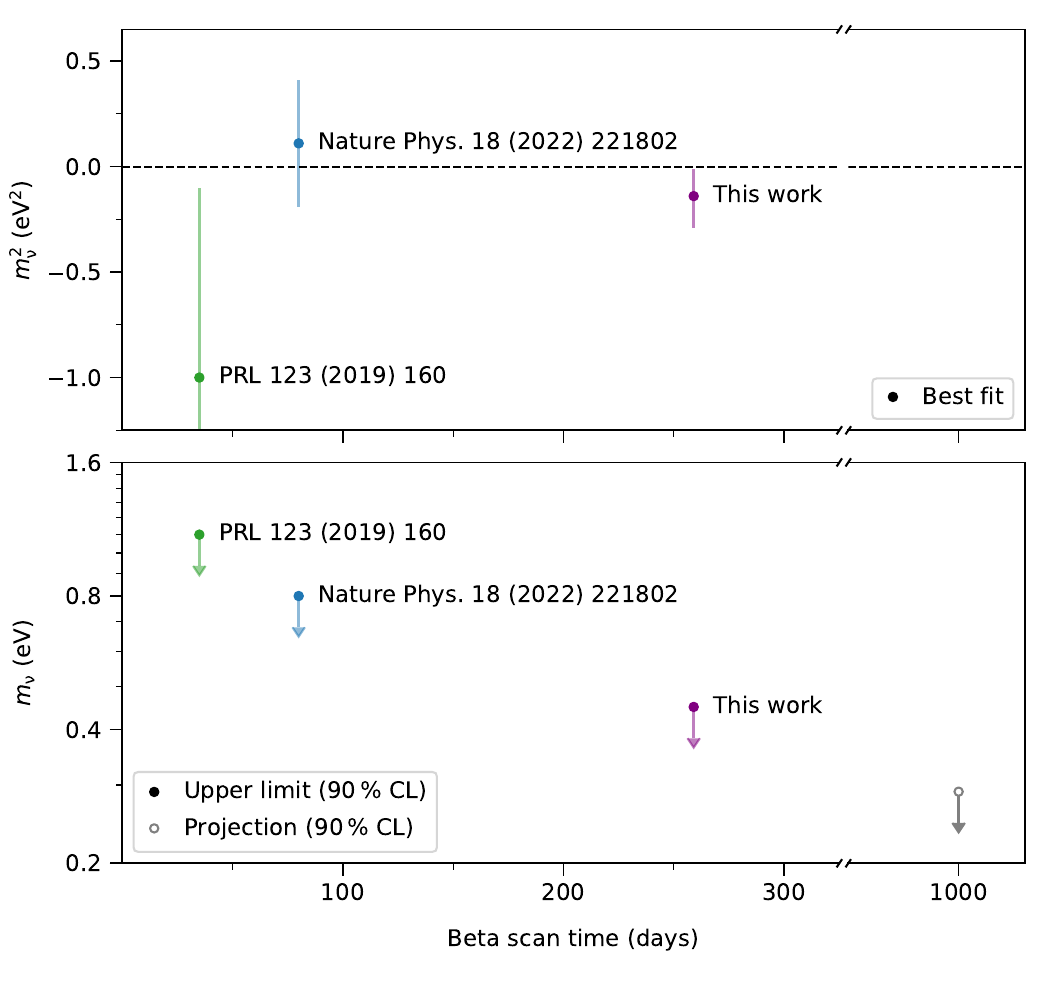}
    \caption{KATRIN neutrino-mass result obtained in this work (five measurement campaigns, purple) compared to previous KATRIN results (first campaign, green, and combined first and second campaigns, blue).}
    \label{Fig:Evolution}
\end{figure}

In this work, we report on the first five measurement campaigns of the KATRIN experiment, performed over \num{259} measurement days from April 2019 to June 2021. With six times the statistics corresponding to \num{36} million electrons, various improvements of the experimental setup, and better control of systematic effects, the KATRIN sensitivity to the effective electron anti-neutrino mass was improved by about a factor of two compared with the previous result~\cite{KATRIN:2021uub}. These improvements include a new spectrometer setting that reduces the background by a factor of two, novel calibration methods using quasi-monoenergetic electrons from the photoelectron source and \nuc{Kr}[83m], and a new approach to assess the molecular final-states uncertainty. We obtain a best-fit value of $\mnu^{2} = \SI[parse-numbers = false]{-0.14^{+0.13}_{-0.15}}{\electronvolt\squared}$, tightening the laboratory-based direct bound to $\mnu <$ \SI{0.45}{\electronvolt} at \SI{90}{\percent} CL. Figure \ref{Fig:Evolution} shows the evolution of the KATRIN results.

KATRIN aims to reach \num{1000} measurement days by the end of 2025. This will correspond to about five times the statistics of the work presented here. 
Based on the current operational conditions, we expect to reach a final sensitivity of better than \SI{0.3}{\electronvolt} at \SI{90}{\percent} CL. Moreover, new methods to further reduce the background~\cite{Gauda:2022vvp} and improve the sensitivity are under investigation.
A new photoelectron source, installed in 2022 at the start of KNM7, allows detailed and high-statistics measurements of the column density and energy-loss function, which are the dominant systematic effects of the current analysis.
The scan-step-duration-dependent background has already been successfully mitigated over the course of the measurements presented here and studies to reduce the uncertainty on the source-potential variations are ongoing.

Beyond the neutrino-mass investigation, the high-statistics and high-precision \be-decay spectrum measured by KATRIN is used to search for physics beyond the Standard Model of elementary particles, such as sterile neutrinos~\cite{KATRIN:2022ith} and Lorentz-invariance violation~\cite{KATRIN:2022qou}, and to probe for the local neutrino overdensity~\cite{KATRIN:2022kkv}.

%% file: 08acknowledgment.tex
\section*{Acknowledgments}

\small
We acknowledge the support of Helmholtz Association (HGF), Ministry for Education and Research BMBF (05A23PMA, 05A23PX2, 05A23VK2, and 05A23WO6), the doctoral school KSETA at KIT, Helmholtz Initiative and Networking Fund (grant agreement W2/W3-118), Max Planck Research Group (MaxPlanck@TUM), and Deutsche Forschungsgemeinschaft DFG (GRK 2149 and SFB-1258 and under Germany's Excellence Strategy EXC 2094 – 390783311) in Germany; Ministry of Education, Youth and Sport (CANAM-LM2015056, LTT19005) in the Czech Republic; Istituto Nazionale di Fisica Nucleare (INFN) in Italy; the National Science, Research and Innovation Fund via the Program Management Unit for Human Resources \& Institutional Development, Research and Innovation (grant B37G660014) in Thailand; and the Department of Energy through Awards DE-FG02-97ER41020, DE-FG02-94ER40818, DE-SC0004036, DE-FG02-97ER41033, DE-FG02-97ER41041, {DE-SC0011091 and DE-SC0019304 and the Federal Prime Agreement DE-AC02-05CH11231} in the United States. This project has received funding from the European Research Council (ERC) under the European Union Horizon 2020 research and innovation programme (grant agreement No. 852845). We thank the computing cluster support at the Institute for Astroparticle Physics at Karlsruhe Institute of Technology, Max Planck Computing and Data Facility (MPCDF), and the National Energy Research Scientific Computing Center (NERSC) at Lawrence Berkeley National Laboratory.
\normalsize

%% file: 10supplement.tex
\section*{Supplementary materials}
\label{Sec:Supplement}

In this work, we report on the third data release of the KATRIN experiment, based on the first five neutrino-mass measurement campaigns, KNM1 to KNM5. The data set has a total statistics of \num{36} million events (counts at the detector within a chosen energy range) in an analysis window of $E_0 - \SI{40}{\electronvolt} \leq qU_i \leq E_0 + \SI{135}{\electronvolt}$. As presented in the main text we find a best-fit value of $\mnu^{2} = \SI[parse-numbers = false]{-0.14^{+0.13}_{-0.15}}{\electronvolt\squared}$. The uncertainty is dominated by the statistical error of $\sigma_{\text{stat.}} =$ \SI{0.108}{\electronvolt\squared}. The total systematic uncertainty amounts to $\sigma_{\text{syst.}} =$ \SI{0.072}{\electronvolt\squared}. Based on the Lokhov-Tkachov method~\cite{LokhovTkachov:2015} we place an upper limit on the effective electron anti-neutrino mass of $\mnu < \SI{0.45}{\electronvolt}$ at \SI{90}{\percent} confidence level (CL). This result improves the previous upper limit, based on the first two KATRIN campaigns, by a factor of \num{1.8}. 

In the supplementary material, we give a comprehensive overview of all analysis steps, from data processing to high-level statistical inference, and information about the data release provided alongside this manuscript. The supplement is structured in the following way: Section~\ref{Sec:Supp-DataProcessingSelectionCombination} details the data processing, selection, and combination. Section~\ref{Sec:Supp-TheoModelAndInputs} provides a detailed description of the theoretical spectrum calculation. The calibration methods are discussed in section~\ref{Sec:Supp-Calibration}. Section~\ref{Sec:Supp-Likelihood} provides an overview of the methods used for the combined fit of the data and the limit inferred for \mnu. The experimental settings and features of the individual campaigns are summarized in section~\ref{Sec:Supp-IndividualCampaigns}. Section~\ref{Sec:Supp-Endpoint} explains the inference of the Q-value. The modifications of the analysis procedure after the initial unblinding are summarized in section \ref{Sec:Supp-PostUB}. Finally, a description of the released data is given in section~\ref{Sec:Supp-DataRelease}.

\input{11sup-data}
\input{12sup-theo}
\input{13sup-calibration}
\input{14sup-likelihood}
\input{15sup-campaigns}
\input{16sup-endpoint}

\input{16_5sup-postunblinding}
\input{17sup-release}

%% file: 11sup-data.tex
\section{Data processing, selection, and combination}
\label{Sec:Supp-DataProcessingSelectionCombination}

At each high voltage (HV) set point $U_i$ of the main spectrometer the electron energies are recorded by the focal-plane detector (FPD). Before hitting the FPD, the electrons are accelerated by the post-acceleration electrode (PAE), which boosts the electron energy, $E$, by $qU_{\text{PAE}} =$ \SI{10}{\kilo\electronvolt}.  The resulting spectrum is governed by the energy resolution of the FPD of about \SI{3}{\kilo\electronvolt} full-width at half maximum, see \cite{Amsbaugh:2014uca} and section~\ref{SubSec:Detector} for details. In order to obtain the count rate at each HV set point, the events in an asymmetric region of interest (ROI) are counted. A ROI of \mbox{$\SI{14}{\kilo\electronvolt} \leq E + qU_{\text{PAE}} \leq \SI{32}{\kilo\electronvolt}$} is used for the KNM1 and KNM2 campaigns, while a narrower ROI of \mbox{$\SI{22}{\kilo\electronvolt} \leq E + qU_{\text{PAE}} \leq \SI{34}{\kilo\electronvolt}$} was selected for the KNM3 to KNM5 campaigns, to reduce the detector background contribution. 

The FPD consists of \num{148} pixels. For the analysis \num{117} and \num{126} so-called golden pixels were chosen for KNM1 to KNM2 and KNM3 to KNM5, respectively. Most of the pixels in the two outermost rings are excluded as structural components of the beamline shadow the flux of electrons at these locations. Pixels with an elevated intrinsic noise are also excluded. 

 \begin{figure}[!t]
     \centering
     \includegraphics[width=0.7\textwidth]{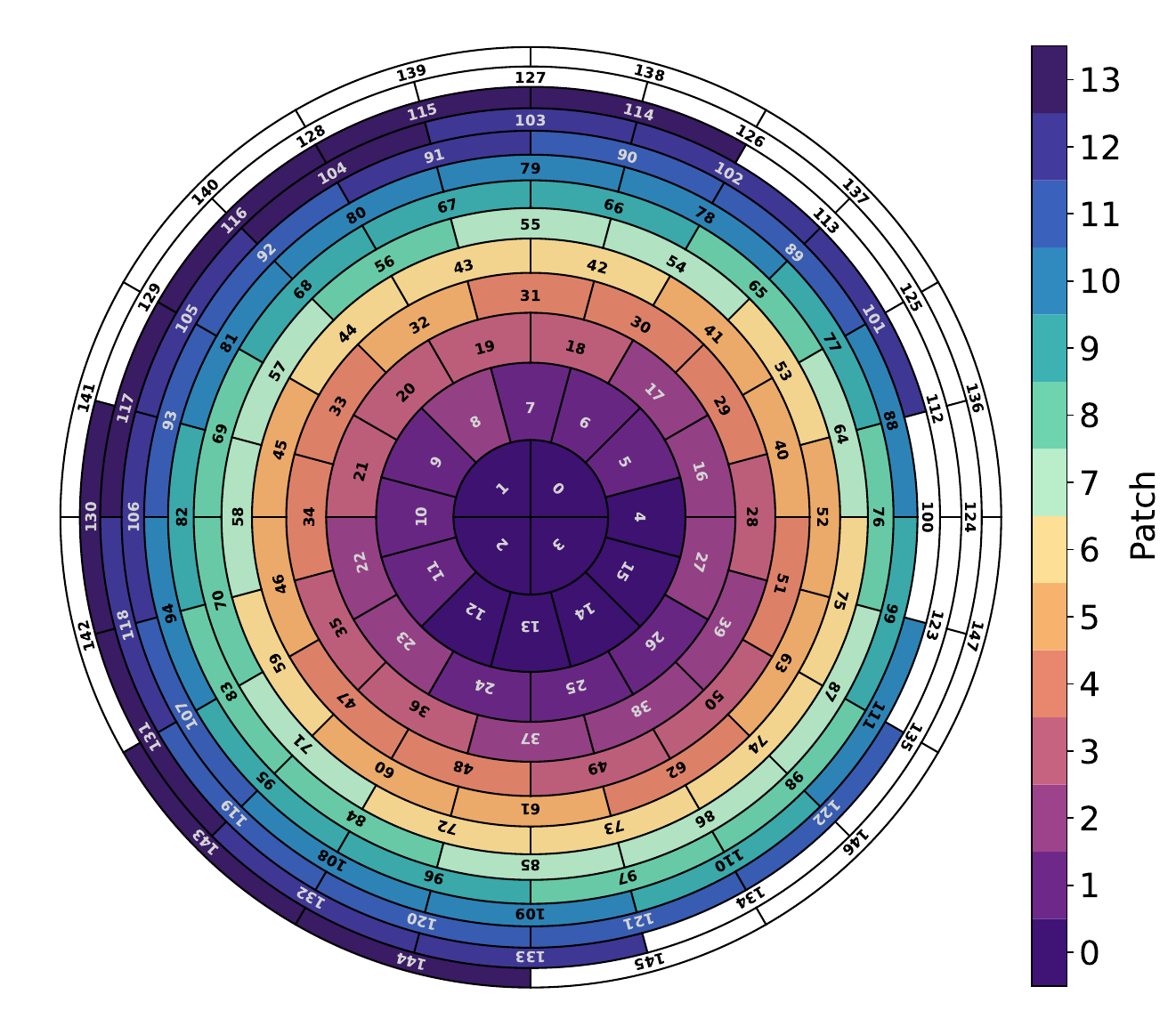}
     \caption{Segmented pixel map of the focal-plane detector with the patch layout of the SAP campaigns. The combination of pixels into patches is defined by the spectrometer transmission properties for electrons arriving at each active colored pixel. White pixels are excluded from the analysis.}
     \label{Fig:Patches}
 \end{figure}
 
For the analysis of the KNM1, KNM2, and KNM3-NAP campaigns the counts in all golden pixels are summed. It is possible to have the same model description for all the pixels, if the nominal analyzing plane (NAP) configuration is used. In the KNM3-SAP, KNM4, and KNM5 campaigns the pixels are grouped into \num{14} so-called patches. On the one hand, grouping of pixels is necessary to reduce the computational cost. On the other hand, segmentation is necessary to take into account that electrons recorded at different detector positions have experienced different retarding potential and magnetic field in the shifted-analyzing-plane (SAP) configuration. The patch definition is chosen to minimize the variation of electromagnetic fields within one patch. Due to a slight misalignment between the detector wafer and the beamline axis of $\mathcal{O}(\SI{1}{\milli\meter})$, the patches do not coincide perfectly with the detector rings, as can be seen in figure~\ref{Fig:Patches}. Each patch contains nine detector pixels, which are selected based on the respective transmission properties.

For each HV set point, the live time starts when the reading of the main spectrometer voltage is within \SI{20}{\milli\volt} of the set point value~\cite{Rodenbeck:2022iys}. Due to the high precision and reproducibility of the HV set points with the advanced post-regulation system~\cite{Rodenbeck:2022iys}, the counts recorded at the same HV set points can be added without introducing any significant spectral distortion. The variance reaches below $\SI{E-3}{\electronvolt^2}$ and is negligible for the \mnusq estimation. Therefore, all the selected scans in a campaign are combined in one single data set, where the number of counts and the live time for each HV set point are summed and the corresponding retarding energies are averaged.

 \begin{figure}[!t]
     \centering
     \includegraphics[width=1.0\textwidth]{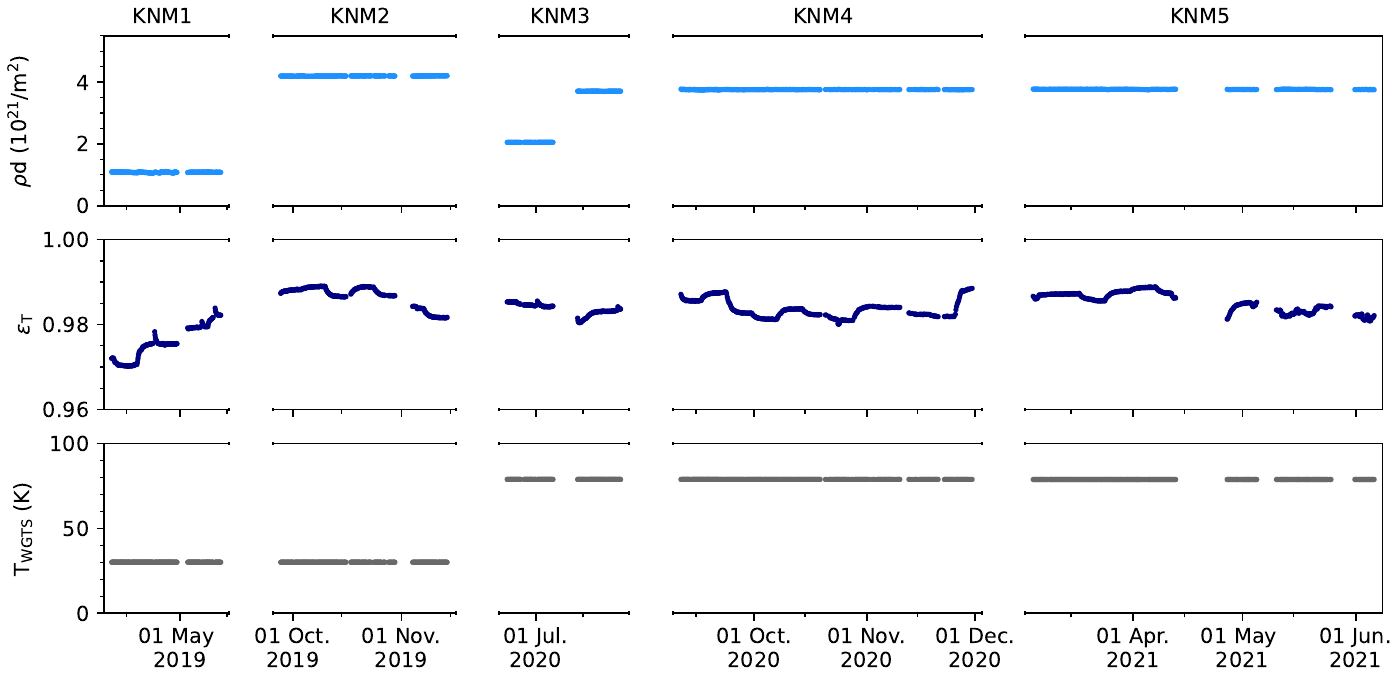}
     \caption{Evolution of source operating parameters throughout the first five measurement campaigns. The column density $\rho d$ (top), tritium purity $\varepsilon_T$ (center), and source temperature $T_\mathrm{WGTS}$ (bottom) are shown. Each data point corresponds to one scan, the uncertainties are too small to be visible.}
     \label{Fig:SlowControl}
 \end{figure}

For each campaign, the so-called golden scans are selected based on tight limits for the key slow-control parameters. Example parameters are shown in figure~\ref{Fig:SlowControl}. A typical reason for rejecting scans is missing readings of key parameters from one of the subsystems, such as the laser Raman system or the high-voltage regulation. Scans can also be discarded due to instabilities observed by monitoring tools, such as spikes in the magnet systems or drifts of the rear-wall bias voltage. The number of golden scans for each measurement campaign is provided in table~1 of the main document.

\begin{figure}[!t]
    \centering
    \includegraphics[width=1.0\textwidth]{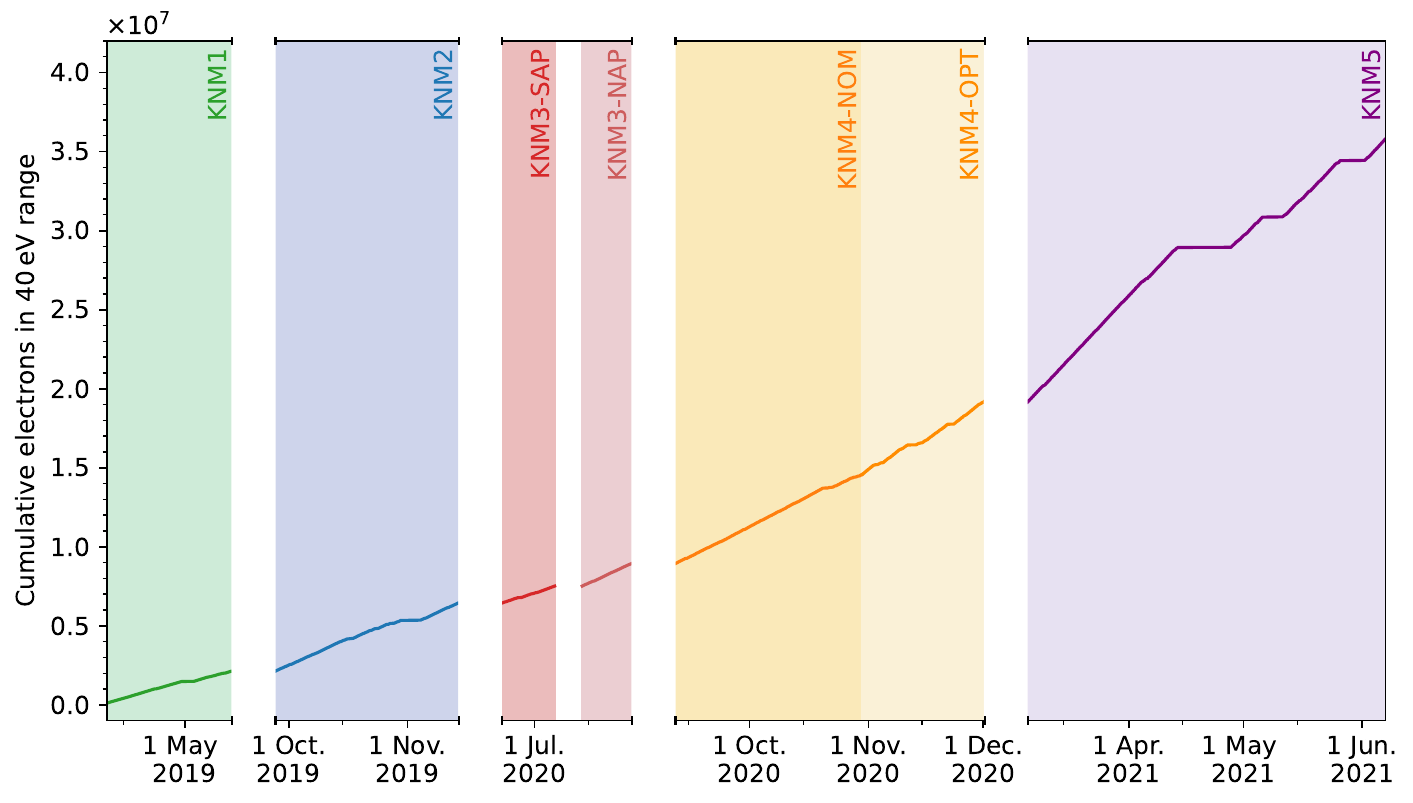}
    \caption{Cumulative counts collected in the $qU>E_0-\SI{40}{\electronvolt}$ analysis window of the first five measurement campaigns. Each campaign is highlighted in the corresponding color.}
    \label{Fig:MeasurementCampaigns}
\end{figure}

Figure~\ref{Fig:MeasurementCampaigns} shows the cumulative counts collected during the neutrino-mass measurements, and only includes the events counted by golden pixels in the analysis range of the golden scans. In total, \num{3.6e7} counts were collected from KNM1 to KNM5. The slopes differ between the measurement campaigns due to changes in the experimental conditions, as detailed in the following:

\begin{itemize}
    \item From KNM1 to KNM2 the column density, and hence the source activity was increased. It was then lowered in KNM3-SAP and raised again to its final value from KNM3-NAP onward, see table~1 of the main document.%
    \item From KNM3-SAP onward the narrow ROI was used. It improves the signal-to-noise ratio but reduces the overall rate, see section~\ref{SubSec:Detector}. 
    \item From KNM4 onward the scan length was increased by \SI{50}{\percent}. The amount of waiting time between HV set points, relative to the total measurement time, is therefore reduced.
    \item KNM5 used an optimized measurement-time distribution (relative time spent at each HV set point), improving the statistical sensitivity on the neutrino mass.
\end{itemize}

A stable operation is essential for a precise measurement of the integral tritium $\upbeta$-decay spectrum. Hence, only small fluctuations of the source parameters are allowed throughout the measurement campaigns. The evolution of the column density, the tritium purity, and the source temperature is shown in figure~\ref{Fig:SlowControl}. The column density, $\rho d$, had a set value of \SI{1.08e21}{\per\metre\squared} in the ``burn-in" KNM1 phase. It was then raised to \SI{4.20e21}{\per\metre\squared} for KNM2. During KNM3, both the new SAP configuration and the new tritium-circulation mode capable of krypton co-circulation were tested. The source temperature, ${T}_{\text{WGTS}}$, was raised from \SI{30.05}{\kelvin} in KNM2 (the ``\SI{30}{\kelvin}'' mode) to \SI{78.85}{\kelvin} in KNM3-SAP (the ``\SI{80}{\kelvin}'' mode). This new operating temperature allows co-circulation of krypton with tritium, resulting in robust calibration and measurements of systematic parameters under the same operating conditions as the neutrino-mass measurements. The column density was set to \SI{2.05e21}{\per\metre\squared} and \SI{3.70e21}{\per\metre\squared} in KNM3-SAP and KNM3-NAP, respectively. During KNM4 and KNM5 the column density was set to its final value of \SIrange[range-units=single]{3.76e21}{3.77e21}{\per\metre\squared}. The tritium purity, $\varepsilon_\mathrm{T}=\left[ N_\mathrm{T_2} + \frac{1}{2}\left( N_\mathrm{HT}+N_\mathrm{DT} \right) \right]/N_\mathrm{tot}$, varies between \num{0.97} and \num{0.99} throughout all measurement campaigns (with $N_\mathrm{T_2}$, $N_\mathrm{HT}$ and $N_\mathrm{DT}$ being the number of the T$_2$, HT, and DT molecules in the source and $N_\mathrm{tot}$ the total number of molecules measured by the Laser Raman system~\cite{LARA_paper2020}). The maximum column density in the ``\SI{80}{\kelvin}'' mode is about \SI{3.8e21}{\per\metre\squared}. The temperature variation in both modes is about \SI{0.1}{\percent}, which fulfills the design requirements.

%% file: 12sup-theo.tex
\section{Theoretical model and inputs} 
\label{Sec:Supp-TheoModelAndInputs}
This section describes the theoretical model and corresponding model inputs. The model is constructed from the differential \be-decay spectrum, integrated over the response function that encodes the working principle of KATRIN.

\subsection{Differential spectrum}

The differential \be-spectrum of molecular tritium, T$_2$, is modeled as a point-like Fermi interaction, which describes the weak decay. Fermi's golden rule 
is used to calculate the differential decay rate~\cite{Otten:2008zz,Kleesiek:2018mel}, 
\begin{equation}
\begin{aligned}
    R_\upbeta(E) = \,\,&
    \frac{G^2_\mathrm{F}|V_\mathrm{ud}|^2}{2\pi^3}|M_\mathrm{nuc}|^2 F(Z,E) p(E+m_e) \\
    &\cdot \sum_i |U_{\mathrm{e}i}|^2 (E_0-E) \\
    &\cdot\sqrt{(E_0-E)^2-m^2_i}~ \Theta(E_0-E-m_i).
    \label{Eq:DiffSpec}
\end{aligned}
\end{equation}
Here $G_\mathrm{F}$ is the Fermi coupling constant, $V_\mathrm{ud}$ is the relevant entry of the quark mixing matrix, $|M_\mathrm{nuc}|$ is the energy-independent nuclear-transition-matrix element of the super-allowed transition, and $\Theta$ is the Heaviside function. The relativistic Fermi function, $F(Z, E)$, describes the Coulomb interaction of the \be-decay electron with the daughter nucleus of charge $Z = 2$. The momentum, $p$, of the electron with kinetic energy $E$, and the neutrino energy $(E_0 - E)$ and momentum $\sqrt{\left(E_0 - E\right)^2 - m^2_i}$ determine the shape of the \be-decay spectrum near the tritium endpoint energy, $E_0$. The endpoint energy is defined as the maximum kinetic energy of the electron assuming zero neutrino mass. The spectrum contains an incoherent sum over the neutrino mass eigenstates, $m_i$ with $i \in \{1, 2, 3\}$, weighted by the squared elements of the neutrino mixing matrix, $|U_{ei}|^2$. In the quasi-degenerate regime, $m_1\approx m_2\approx m_3$, this sum is replaced by the effective squared neutrino mass $\mnu^2 = \sum_i |U_{ei}|^2 m_i^2$.

The released energy, $Q$, is split between the electron, the neutrino, and the daughter molecular ion. The daughter ion acquires energy in terms of recoil and rotational, vibrational, and electronic excitation states. The total energy available for the neutrino is $\varepsilon_f = E_0 - V_f - E$, where $V_f$ corresponds to the excitation energy of a given state $f$. $V_f$ and the corresponding transition probabilities $P_f$ are described by the so-called final-states distribution (FSD). The resulting differential energy spectrum is given by
\begin{equation}
\begin{aligned}
    R_\upbeta(E;E_0,m_\upnu^2) = \,\,& \frac{G^2_\mathrm{F}|V_\mathrm{ud}|^2}{2\pi^3}|M_\mathrm{nuc}|^2 F(Z,E) p(E+m_e) \\
    &\cdot \sum_f P_f G(E, E_0-V_f)\varepsilon_f \\
    &\cdot \sqrt{\varepsilon_f^2-m^2_\upnu} \Theta(\varepsilon_f-m_\upnu).
    \label{Eq:DiffSpecT2}
\end{aligned}
\end{equation}

The additional radiative correction factor $G(E, E_0 - V_f)$ accounts for higher-order quantum-electrodynamics contributions \cite{PhysRevC.28.2433}. 
Other theoretical corrections are negligible and not included in the analysis.

\subsection{Final-states distribution}

The FSD, $({V_f, P_f})$, is a key input to evaluate equation~\ref{Eq:DiffSpecT2}. An ab initio calculation of the FSD was performed assuming the sudden approximation, which neglects the Coulomb interaction of the \be-electron with the remaining molecular system, $^3$HeT$^+$, \cite{Saenz:2000dul,Doss:2006zv}. Note that the leading interaction is included in the Fermi function in eq.~\ref{Eq:DiffSpecT2}.

\begin{figure}[!t]
     \centering
     \includegraphics[width=0.8\textwidth]{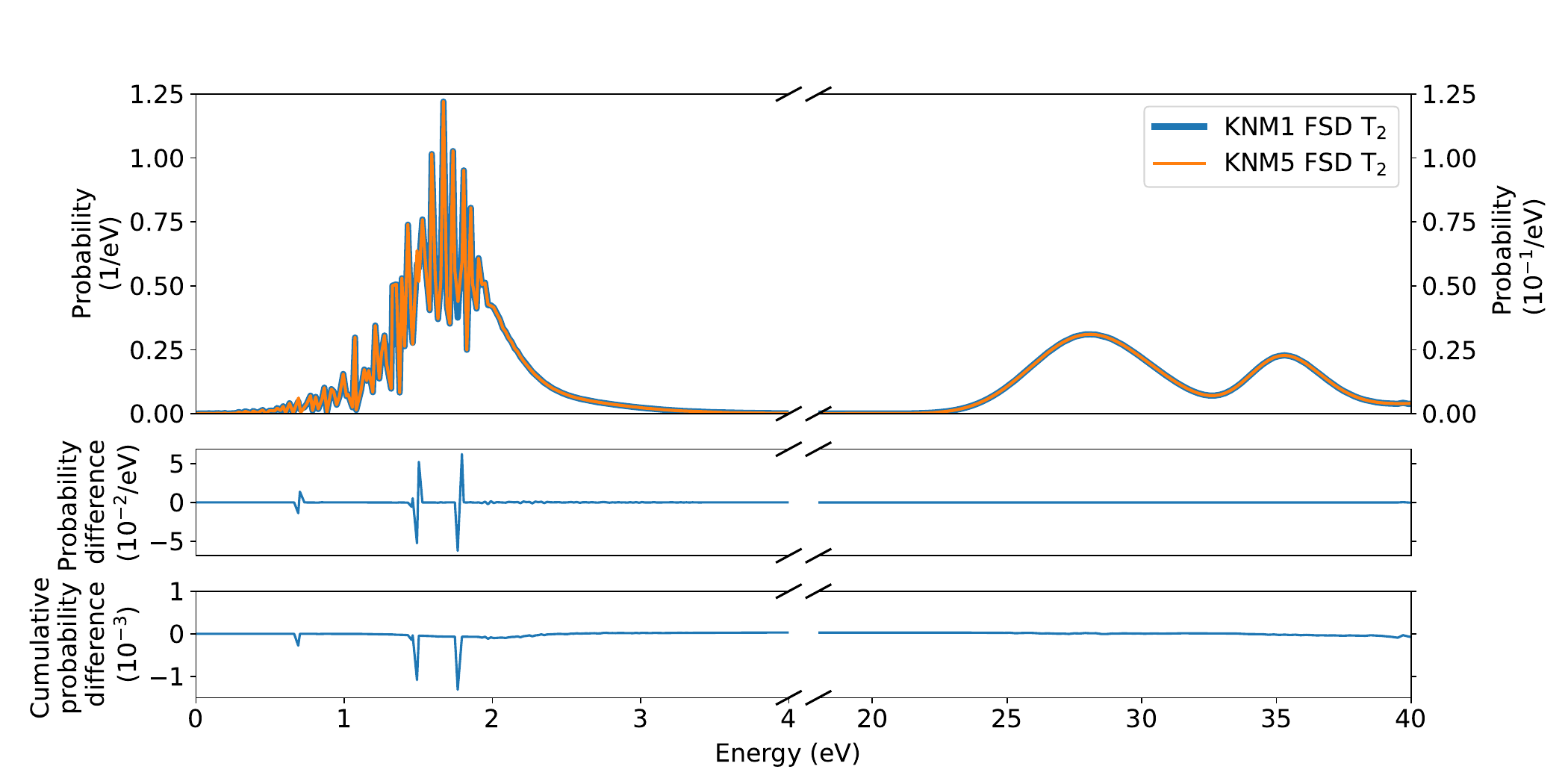}
     \caption{Molecular final-states distribution for T$_2$, evaluated at a temperature of \SI{30}{\kelvin}. The top panel shows the FSD used in the previous analyses of KNM1 and KNM2 (``KNM1 FSD'', blue) \cite{KATRIN:2019yun,KATRIN:2021uub} as well as the FSD obtained with the improved calculation (``KNM5 FSD'', orange)\cite{FSDNewPaper}. The part below \SI{4}{\electronvolt} describes transitions into the electronic ground state of $^3$HeT$^+$, while the higher energies describe electronically excited states and the dissociation continuum. To emphasize the small discrepancy between the two almost overlapping FSDs, the absolute difference is shown in the middle panel. The main discrepancies are related to binning effects and are not relevant after summation over all final states, as shown in the bottom panel, which depicts the difference of the cumulative probability densities.}
     \label{Fig:FSD}
 \end{figure}

The FSD was re-calculated for this analysis. The difference between the FSD used for the previous KATRIN analyses (the ``KNM1 FSD'') and the new ``KNM5 FSD'' is shown in figure~\ref{Fig:FSD}. This update has negligible impact on the neutrino-mass estimation. However, the re-calculated FSD allows systematic studies of FSD-related uncertainties. A dedicated procedure to estimate the impact of FSD-related uncertainties on the neutrino-mass inference was developed~\cite{FSDNewPaper}. 
The previous estimates were based on direct modifications of the FSD, allowing $\mathcal{O}(\SI{1}{\percent})$ variations of the variance of the distribution and of the ratio of the ground state and excited state probabilities~\cite{KATRIN:2021fgc}. 
In contrast to these conservative and general estimates, the new procedure examines specific effects that could impact the FSD. The effects are considered at the stage of FSD computation and include the convergence of the calculation, the use of external inputs, and the validity of approximations and corrections in the calculation. An FSD, calculated including the above-mentioned effects, is used to fit a simulated benchmark data set, and the resulting bias of the $\mnu^2$ parameter is considered as an estimate of the additional $\mnu^2$ systematic uncertainty. The largest impact on $\mnu^2$ comes from the adiabatic, relativistic, and radiative corrections and amounts to less than $\SI{0.001}{\electronvolt^2}$ bias. Systematic contributions of this order are neglected in the presented analysis, see table~2 of the main document.

The FSDs for three isotopologues (T$_2$, DT, and HT) are calculated for different initial quantum states of molecular angular momentum, which are then weighted according to the population of the states following a Boltzmann distribution for the given source temperature.
To facilitate the spectrum computation, all effects that can be described as an additional Gaussian broadening -- for example, Doppler broadening or the source and the spectrometer potential variations -- are emulated as a convolution of the FSD with the corresponding distribution.
As part of the blinding procedure, the FSDs are broadened by an unknown value, so that the true fit result is not accessible during analysis preparations.

\subsection{Response function}

The experimental response function,
\begin{align}
    f_\mathrm{calc}(E,qU) = & \int_{\epsilon=0}^{E-qU}
    \int_{\theta=0}^{\theta_\mathrm{max}} \mathcal{T}(E-\epsilon,\theta,qU)\sin\theta \cdot \sum_s P_s(E) f_s(\epsilon) \mathrm{d}\theta \mathrm{d}\epsilon,
    \label{Eq:ResponseFunction}
\end{align}
describes the probability of a \be-decay electron emitted in the source with kinetic energy $E$, to reach the FPD after propagating through the source and the main spectrometer. Transmitted electrons are selected by the transmission condition of the spectrometer, $\mathcal{T}(E - \epsilon,\theta,qU)=1$, while those electrons reflected back to the source by the magnetic field and the electrostatic field have $\mathcal{T}(E - \epsilon,\theta,qU)=0$. $\mathcal{T}$ depends on the kinetic energy of the electron after energy loss $\epsilon$ from inelastic scattering off the tritium molecules in the source; the retarding energy $qU$, where $q$ is the charge of the electron and $U$ is the voltage applied to the main spectrometer; and the pitch angle $\theta$ between the momentum of the electron and the magnetic field in the source. $\mathcal{T}$ depends also on the magnetic fields at the starting position of the electron, $B_\mathrm{src}$, the minimal magnetic field in the main spectrometer, $B_\mathrm{ana}$, and the maximal magnetic field, $B_\mathrm{max}$. The maximal acceptance angle of the electrons is defined by the ratio of the magnetic fields, $\theta_\mathrm{max}=\arcsin{\sqrt{B_\mathrm{src}/B_\mathrm{max}}}$. For more details see~\cite{Kleesiek:2018mel}.

\begin{figure}[!t]
     \centering
     \includegraphics[width=0.5\textwidth]{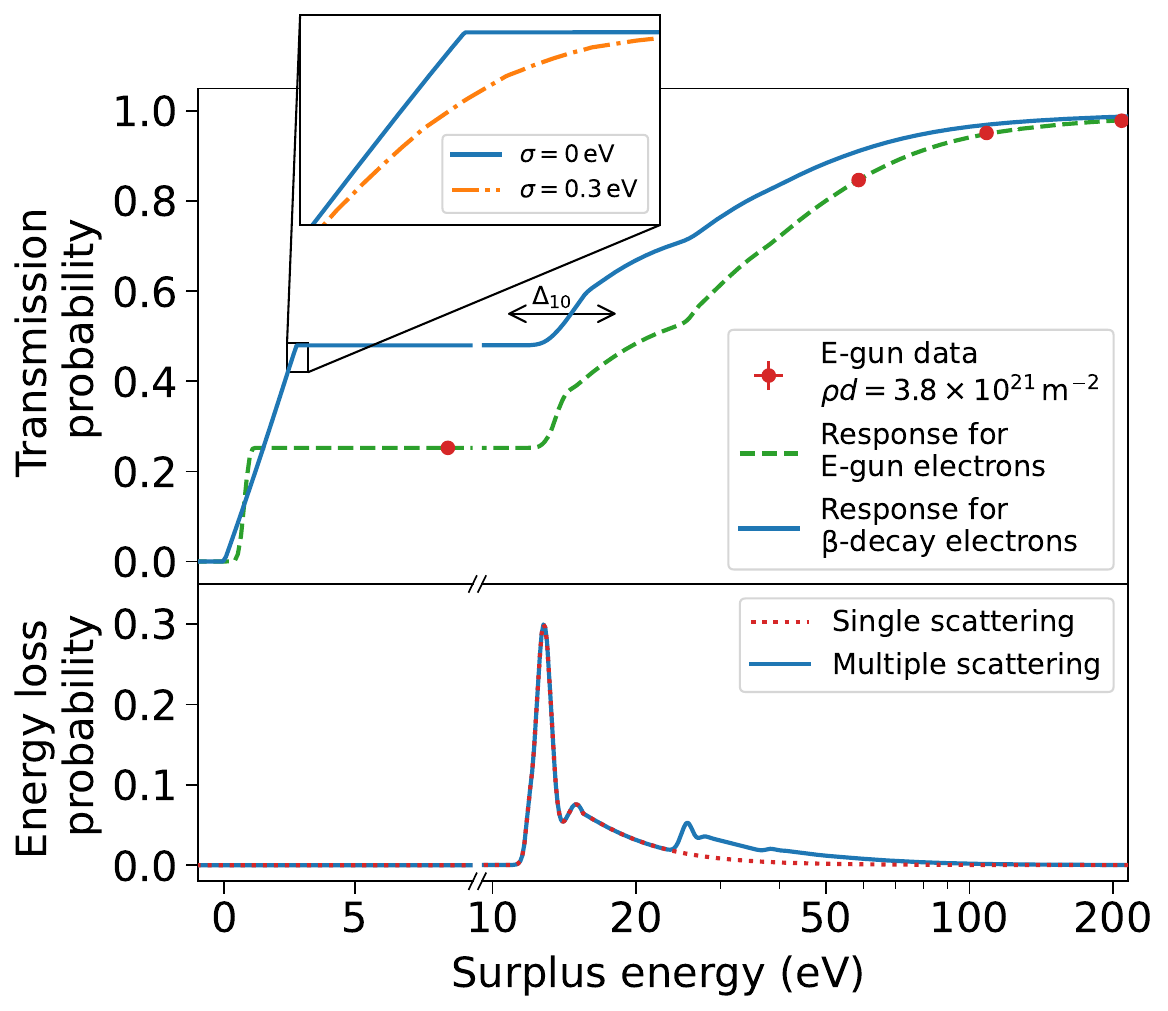}
     \caption{An illustration of the response function, the measurement of the column density in the tritium source, and the impact of the source potential (top). Note the change from linear to log-scale at \SI{9.5}{\electronvolt} in both plots, indicated by the energy-axis break. The relative electron-gun rate (red circles) is measured at four surplus energies, defined as the electron kinetic energy above the retardation potential. The relative uncertainties of the electron-gun rate are not visible in the plot. A fit of the response function to the data (green, dashed line) yields a column density of $\rho d =$ \SI{3.8e21}{\per\meter\squared} for this specific measurement (the response function description is found in section~\ref{Sec:Supp-TheoModelAndInputs}, equation~\ref{Eq:ResponseFunction}). The solid blue line illustrates a calculated response function of the \be-decay electrons with an isotropic pitch-angle distribution up to the maximum acceptance angle. The \SI[parse-numbers=false]{2.7}{\electronvolt} width of the transmission edge is characteristic for the magnetic field of about $\SI{0.6}{\milli\tesla}$ in the analyzing plane and the maximum magnetic field of \SI{4.2}{\tesla}. As an approximation, the source potential can be modeled by a transmission broadening (shown in the inset with an exaggerated value \SI{0.3}{\electronvolt}, a typical broadening is 10-fold smaller) and a shift $\Delta_{10}$ of the energy loss region of the response. The bottom panel shows the energy-loss function for a single scattering process (red-dotted line) and multiple scatterings (blue line) as they occur for the measured $\rho d$ value (for details see\cite{KATRIN:2021rqj}).} 
     \label{Fig:Response}
 \end{figure}

The probability for an electron to undergo $s$ scatterings in the source is given by $P_s(E)$, while the probability to lose energy $\epsilon$ after $s$-fold scattering is given by the energy-loss function $f_s(\epsilon)$. The energy-loss function was measured in situ with the monoenergetic angular-selective source of photoelectrons (electron gun) installed at the upstream end of the setup~\cite{KATRIN:2021rqj}, see lower panel in figure~\ref{Fig:Response}. The scattering probabilities $P_s(E)$ for a certain angle and starting position are approximated by a Poisson distribution depending on the column density and the energy-dependent scattering cross section~\cite{Liu_PhysRevA.35.591}. Electrons starting with different pitch angles take different paths through the source; therefore, the scattering probabilities also depend on the angle $\theta$. This effect is taken into account for $s \in \{0,1\}$ but is negligible for $s\ge2$. A typical response function is shown in the upper panel of figure~\ref{Fig:Response}. 

While moving through the strong magnetic field of the source and the transport section, a small part of the electron energy of up to $\mathcal{O}(\SI{100}{\milli\electronvolt})$ is lost in the form of synchrotron radiation. To account for this energy loss, which depends on the electron pitch angle, the maximal acceptance angle $\theta_\mathrm{max}$ is modified in the integration of equation~\ref{Eq:ResponseFunction}. The updated maximum angle is obtained numerically, solving the equation $E_\mathrm{surplus}(\theta^\prime_\mathrm{max}) = 0$, where the available surplus energy $E_\mathrm{surplus}=E-qU$ in the analyzing plane is reduced by the synchrotron energy loss. For details see~\cite{Kleesiek:2018mel}.

An additional effect is related to the angular-dependent detection efficiency of the FPD. Electrons exiting the spectrometer are de-collimated by the high magnetic field in the detector region, $B_\mathrm{det} \approx B_\mathrm{src} =$ \SI{2.5}{\tesla}. Electrons with higher angles have a lower probability to be counted in the ROI (see section \ref{SubSec:Detector}). Therefore, the relative detection efficiency $\varepsilon_\mathrm{eff}$ depends on the angle $\theta$, which can be described by a phenomenological model $\varepsilon_\mathrm{eff}(\theta) = c_0+c_1\cdot \theta + c_2\cdot \theta^2$. The coefficients $c_{0,1,2}$ are obtained by simulation. The factor $\varepsilon_\mathrm{eff}(\theta)$ is added to the calculation of equation~\ref{Eq:ResponseFunction}.

\subsection{Integrated spectrum}

The measured rate of \be-decay electrons for a given retarding energy set point $qU_i$ is described by a convolution of the differential spectrum rate, equation~\ref{Eq:DiffSpecT2}, with the experimental response, equation~\ref{Eq:ResponseFunction},
\begin{equation}
    R_{\mathrm{int},\upbeta}\left(qU_i\right) = N_{\mathrm{T}} \int_{qU_i}^{E_0}R_\upbeta\left(E; E_0, \mnu^2\right) f_\mathrm{calc}\left(E, qU_i\right) \ \mathrm{d} E.
    \label{Eq:SignalRate}
\end{equation}
The integrated signal rate is proportional to the number of tritium atoms in the source, the maximum acceptance angle, and the absolute detection efficiency, amounting to the signal normalization, $N_{\mathrm{T}}$.

To describe the overall background rate, different contributions are added, 
\begin{equation}
    R_{\mathrm{bg}}\left(qU_i\right) =  R_{\mathrm{bg}}+ R_{\mathrm{spec,det}}\left(qU_i\right)+R_{\mathrm{RW}}\left(qU_i\right).
    \label{Eq:BackgroundRate}
\end{equation}
In addition to a retarding-energy-independent rate $R_\mathrm{bg}$, smaller $qU$-dependent systematic contributions from the spectrometer and detector background rates $R_{\mathrm{spec,det}}\left(qU_i\right)$, are included. The spectrum of \be-decay electrons from the rear wall is obtained by dedicated measurements and is included into the background rate $R_{\mathrm{RW}}\left(qU_i\right)$, taking into account that the rear-wall electrons have to pass the whole tritium gas column to reach the spectrometer. Further details on the spectrometer and detector background as well as rear-wall contributions can be found in section~\ref{SubSec:BackgroundSystematics} and section~\ref{SubSec:RWSpectrum}, respectively.

The resulting model of the integrated spectrum is given by
\begin{equation}
     R_{\mathrm{calc}}\left(qU_i;A, R_\mathrm{bg}, E_0, m_\upnu^2\right) = A\cdot R_{\mathrm{int},\upbeta}\left(qU_i; E_0, m_\upnu^2\right) + R_{\mathrm{bg}}+ R_{\mathrm{spec,det}}\left(qU_i\right)+R_{\mathrm{RW}}\left(qU_i\right).
    \label{Eq:TotalSpectrum}
\end{equation}
The free parameters of the model are the signal normalization factor $A$, the effective endpoint energy $E_0$, the background rate $R_\mathrm{bg}$, and the squared effective neutrino mass $\mnu^2$. The model also depends on multiple systematic parameters that are constrained by calibration measurements and the slow-control monitoring systems, or obtained by simulations. A full list of the systematic parameters is provided in section~\ref{SubSec:ReleaseInputs}.

%% file: 13sup-calibration.tex
\section{Calibration measurements and systematic inputs}
\label{Sec:Supp-Calibration}

This section describes the calibration measurements and simulations that were performed to determine the input values and uncertainties to constrain systematic effects in the neutrino-mass analysis.

\subsection{Column density} 
\label{SubSec:ColumnDensity}

The amount of tritium gas in the source defines the rate of \be-decay processes and the scattering probability of electrons propagating through the source. The probability depends on the column-density parameter $\rho d$, where $\rho$ is the average gas density and $d =$ \SI{10}{\meter} is the total length of the source.

The column density is measured using the electron gun, which emits electrons at a constant rate with well-defined energy and angle. These electrons pass through the entire source and can lose part of their energy through scattering. The product of the column density and the inelastic scattering cross section, $\rho d\times\sigma_\mathrm{inel}$, defines the probability for electrons to undergo $s$ scatterings, while the energy-loss function, $f_s(\epsilon)$, defines the observed shape. An example is shown in figure~\ref{Fig:Response}.

The main systematic uncertainty of the $\rho d\times\sigma_\mathrm{inel}$ determination is connected to the calibration method itself. The path length of the electrons $d^\prime = d/\cos{\theta_\mathrm{egun}}$ is defined by the length of the source, $d$, and the electron angle with respect to the source axis, $\theta_\mathrm{egun}$. After the replacement of the electron gun in 2022, the accuracy of the $\theta_\mathrm{egun}$ determination for the KNM1-5 campaigns was revisited. The angle was then determined retrospectively from the shift of the measured energy spectrum of the emitted electrons at different magnetic field settings in the main spectrometer. An additional correction of $\Delta\theta_\mathrm{egun}=\SI{2.6}{\degree}$ was included based on a newly discovered dependency of the electron angle on the starting energy; a conservative uncertainty of \SI{100}{\percent} of the correction was added. Another effect is related to the electron-gun background, which typically arrives in bunches. This time-correlated behavior is taken into account by an accurate subtraction of the background, based on the FPD spectrum of high-multiplicity events. 

However, the calibration measurements do not allow one to monitor short-term variations of the column density. Its relative evolution is obtained using dedicated sensors, such as pressure sensors or flow meters used in KNM2 and KNM3, or via the tritium spectrum rate that scales with the column density. The latter method was implemented for KNM1, KNM4, and KNM5. The high-rate set points at \SI{300}{\electronvolt} and \SI{90}{\electronvolt} below $E_0$ are used to achieve a small statistical uncertainty for each individual scan. The column density is extracted by fits that exclude the set points in the neutrino-mass analysis range from $E_0 - \SI{40}{\electronvolt}$ to $E_0$, which avoids double use of the neutrino-mass-sensitive data. 

The combination of the electron gun-calibration measurements and the relative evolution provides the $\rho d\times\sigma_\mathrm{inel}$ parameter for each campaign. The column-density uncertainties are between \SIrange{0.92}{1.35}{\percent} and are partially correlated between campaigns. The corresponding covariance matrix can be found in the systematic inputs described in section~\ref{SubSec:ReleaseInputs}.

\subsection{Shifted analyzing plane}
\label{Subsec:Supp-SAP}

\begin{figure}[!t]
     \centering
     \includegraphics[width=0.49\textwidth]{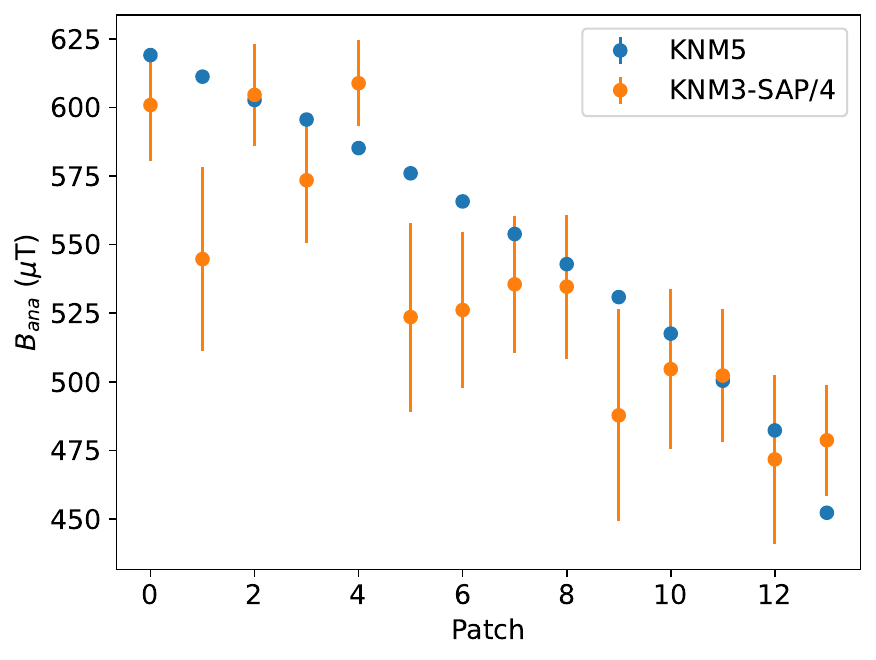}
     \includegraphics[width=0.49\textwidth]{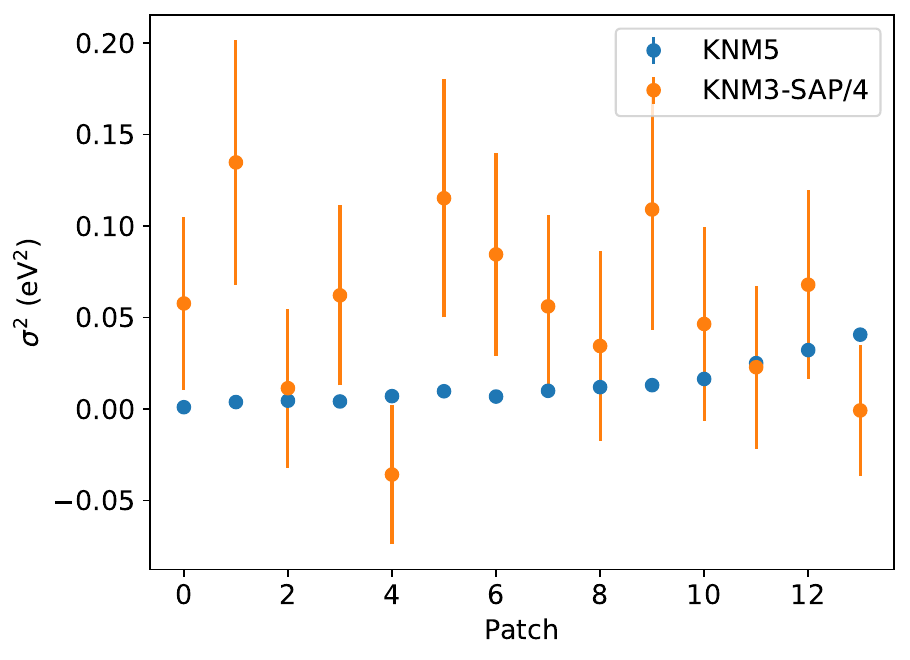}
     \includegraphics[width=0.49\textwidth]{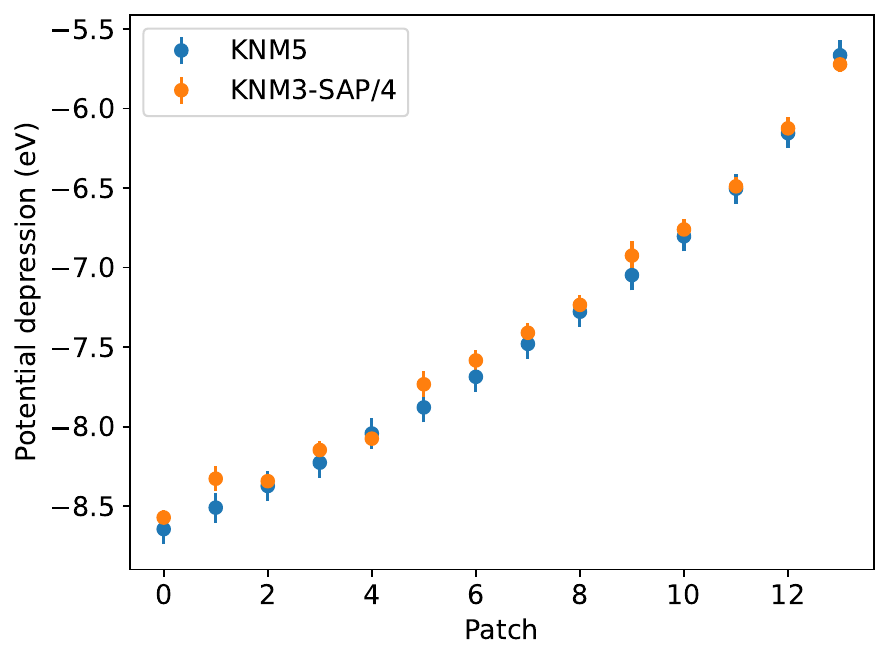}
     \includegraphics[width=0.49\textwidth]{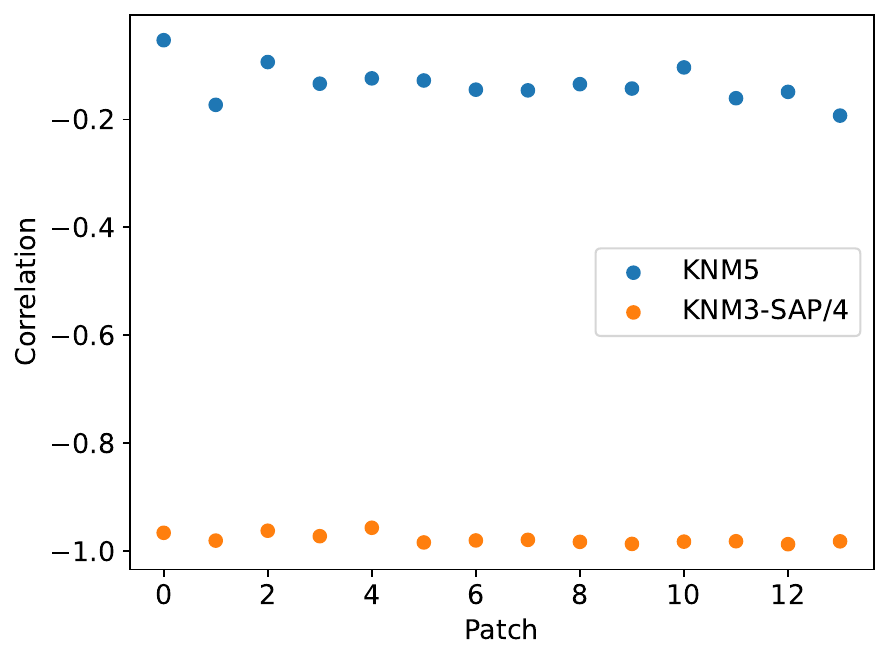}
     \caption{Transmission properties of the shifted-analyzing-plane configuration for the KNM3-SAP, KNM4, and KNM5 campaigns. The magnetic field (top left) and the transmission broadening (top right) have larger uncertainties for the KNM3-SAP and KNM4 campaigns, because the corresponding calibration measurement was performed using the K-32 line with a significant natural line width. The correlation between the magnetic field and the squared broadening (bottom right) becomes small for the KNM5 configuration where the fields were measured using the narrow N$_{2,3}$-32 lines. The potential depression (bottom left) demonstrates the significant variation of the electric potential in the analyzing plane of the SAP configuration.}
     \label{Fig:SAP}
 \end{figure}

The statistical sensitivity to the neutrino mass depends critically on the background rate. The main contribution to the background comes from electrons produced in the main spectrometer volume and accelerated towards the detector by the electric field~\cite{KATRIN:2020zld}. It was shown that these electrons are almost uniformly distributed within the main spectrometer volume. The SAP configuration was designed and proven to reduce the effective volume of the magnetic flux in the main spectrometer while preserving good energy resolution~\cite{Lokhov:2022iag}. The highest retarding potential and the minimum magnetic field are shifted towards the detector, which reduces the spectrometer background by a factor of two. However, this field configuration is highly inhomogeneous;  the variation of electric retarding potential reaches \SI{3}{\volt} and the magnetic field varies by \SI{175}{\micro\tesla}, see figure~\ref{Fig:SAP}. 

In this configuration, the field simulations that were used for the KNM1 and KNM2 analyses become less reliable, as the simulation results depend on possible unknown misalignments in the setup. Therefore, two dedicated in-situ measurements of the electromagnetic fields in the analyzing plane were performed. These measurements used \nuc{Kr}[83m] conversion electrons with a well-defined energy spectrum to infer the transmission properties of the SAP configuration. For the SAP configuration in KNM3-SAP and KNM4, the spectrum of the \SI{17.8}{\kilo\electronvolt} K-32 line with a natural line width of \SI{2.7}{\electronvolt} was scanned. In this measurement, the tritium gas was not injected in the source. After KNM4, the field configuration changed due to a modification of one of the solenoids around the main spectrometer. The fields in this modified configuration were measured using the N$_{2,3}$-32 lines, which have energies of \SI{32.1}{\kilo\electronvolt}, but a negligible natural line width, as depicted in figure~\ref{Fig:Krypton}. The N$_{2,3}$-32 lines measurement was performed with a co-circulation of krypton and tritium at a column density of about \SI{2e21}{\per\meter\squared}, which provides very high rates of conversion electrons but cannot be used for the K-32 line scans due to the high tritium \be-decay rate. The measured parameters are the electric potential $qU_\mathrm{ana}$, or the so-called potential depression, which is the difference between the voltage on the spectrometer and the actual potential in the analyzing plane, the magnetic field $B_\mathrm{ana}$, and the additional variance $\sigma^2$ of the transmission function due to the inhomogeneity of the fields. Source-related effects, such as additional broadenings or shifts, were determined by reference scans of the same lines in a symmetric field configuration with $B_\mathrm{ana} =$ \SI{270}{\micro\tesla}.

Figure~\ref{Fig:SAP} shows the results of the two calibration measurements. The K-32 line analysis is performed using the same detector patch segmentation as the neutrino-mass analysis. The larger uncertainties of the magnetic field strength and transmission function variance are related to the natural line width. The strong anti-correlation between the magnetic field and broadening values is propagated into the neutrino mass analysis, see section~\ref{SubSec:SysPropagation}. The measurement with the N$_{2,3}$-32 lines provides more accurate fit results. The magnetic field and variance are not correlated due to the negligible natural line width. Figure~\ref{Fig:Krypton} shows the spectrum and highlights the impact of the two different magnetic-field settings with $B_\mathrm{ana}$ of \SI{270}{\micro\tesla} and \SI{600}{\micro\tesla}. The effect of the transmission broadening is shown for an exaggerated broadening of $\sigma =$ \SI{0.3}{\electronvolt} in comparison to a spectrum without broadening. The analysis of the N$_{2,3}$-32 scans is performed individually for each pixel and the results are combined into the transmission properties of each patch. Due to the higher energy and applied voltages in the N$_{2,3}$-32 line scans, a small correction of $\mathcal{O}(\SI{50}{\milli\electronvolt})$ obtained through simulations is applied to the potential depression, to recover the values that can be used at the tritium endpoint.

\subsection{Source potential}
\label{SubSec:Plasma}

In equation~\ref{Eq:SignalRate}, the retarding energy $qU$ is formally defined by the voltage $U$ applied to the spectrometer. However, this implicitly assumes that the electric potential at the position of the \be-decay is equal to the ground potential. A non-zero and non-uniform source potential can change the electron starting energy and modify the measured spectrum. 

An offset shifts the effective endpoint, $E_0$, and biases the Q-value determination (see section~\ref{Sec:Supp-Endpoint}), but has no impact on the neutrino mass as the offset does not alter the spectral shape. Spatial and temporal variations of the source potential produce an additional broadening, which is included in the analysis for an unbiased neutrino-mass determination. Longitudinal variations introduce an additional shape distortion. In a simplified view, the electrons, which are scattered once, mostly come from the upstream part of the source. The unscattered electrons are mostly produced in the downstream part, closer to the spectrometer. If the front and the rear parts of the source have different electric potentials, the unscattered and one-time scattered electrons will have different starting energies on average. This difference leads to an effective shift $\Delta_{10}$ of the energies of one-time-scattered electrons in the response function, see figure~\ref{Fig:Response} for illustration. Such a shift, if not taken into account, would cause an additional bias on the squared neutrino-mass parameter. The same applies to multiple-scattered electrons, though the impact on the neutrino mass gets smaller with the increasing number of scatterings.

\begin{figure}[!t]
     \centering
     \includegraphics[width=0.8\textwidth]{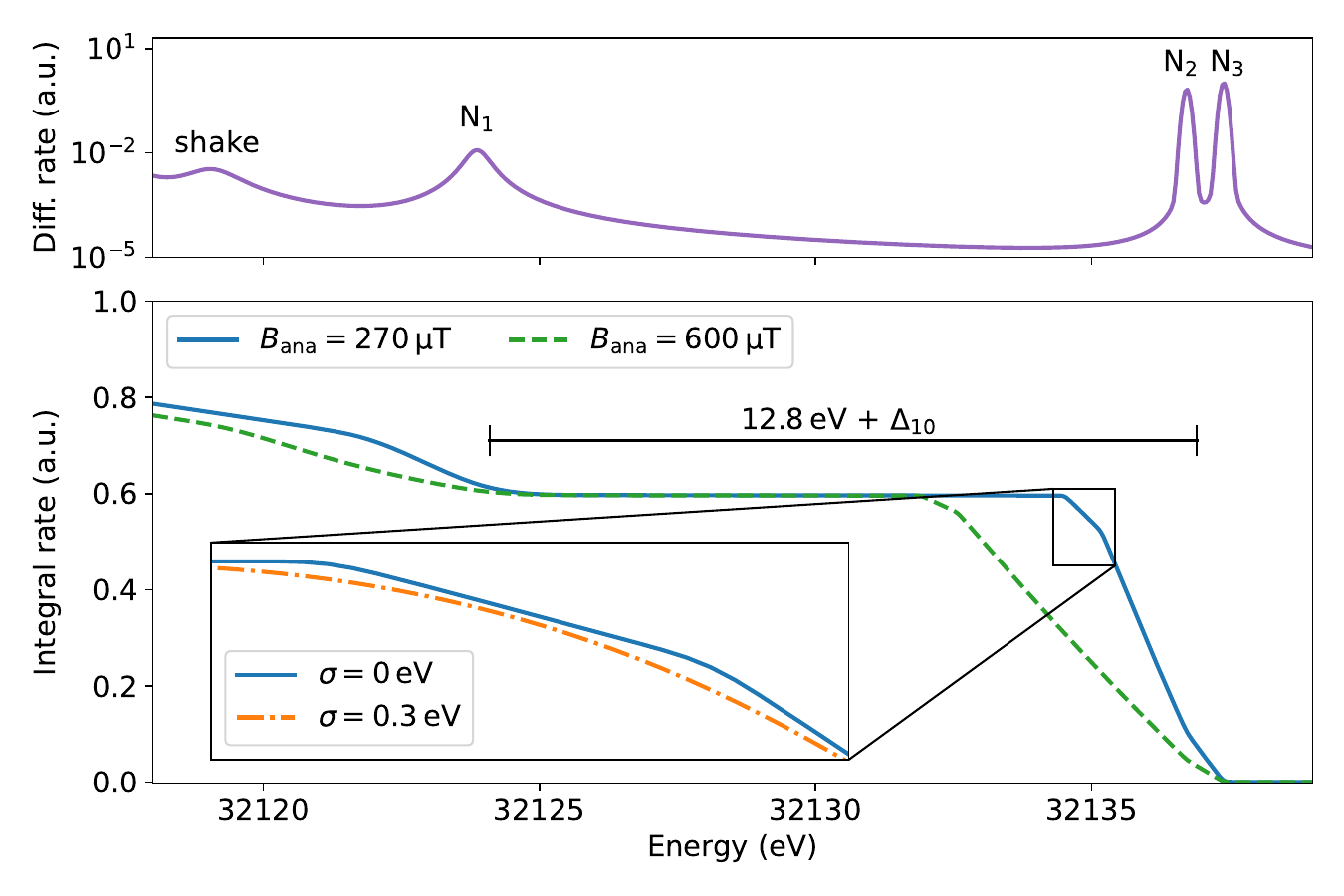}
     \caption{Models of the differential (top) and integrated (bottom) \nuc{Kr}[83m] conversion electron spectra. The differential spectrum is broadened by the Doppler effect for the source temperature of \SI{80}{\kelvin}. The integrated spectrum, solid (blue) line, highlights the $\Delta_{10}$ parameter, associated with longitudinal variations of the source potential. This effective parameter describes an additional shift of the one-time-scattered electrons, which appears in the spectrum at approximately \SI{12.8}{\electronvolt} below the main line. The comparison of the dashed (green) line and the solid (blue) line shows the impact of different magnetic fields in the analyzing plane of the main spectrometer, \SI{600}{\micro\tesla} and \SI{270}{\micro\tesla} respectively. The dash-dotted (orange) line shows the impact of an additional hypothetical energy scale broadening, $\sigma =$ \SI{0.3}{\electronvolt}, on the integrated spectrum. The effective line position is defined by the retarding potential in the main spectrometer.}
     \label{Fig:Krypton}
 \end{figure}

The variations of the source potential are measured with \nuc{Kr}[83m] conversion electrons, using the new operation mode of the tritium loops, which allows co-circulation of krypton and tritium in the same conditions as in the \be-spectrum scanning mode~\cite{KATRIN:2022zqa}. The rear-wall bias voltage was chosen to provide the best radial homogeneity of the source potential, measured with the position of L$_3$-32 line. The variation of the source electric potential is measured with the N$_{2,3}$-32 lines. These lines have negligible natural line width and allow measurement of the broadening with high accuracy. Longitudinal variations lead to an effective shift of the one-time-scattered electrons. This shift, $\Delta_{10}$, is illustrated in figure~\ref{Fig:Krypton}. 

A systematic uncertainty of the determination of $\Delta_{10}$ is related to the N$_1$-32 line, which appears at approximately the same position as the one-time-scattered N$_{2,3}$-32 electrons. Another source of uncertainty is the energy-loss function, which was measured for the electrons of \SI{18.6}{\kilo\electronvolt} energy but not for \SI{32.1}{\kilo\electronvolt} electrons. 
These uncertainties are under investigation and the measured value of $\Delta_{10}$ is not used. Instead, we use a constraint from an ab initio consideration.
Any longitudinal variation of the source potential causes a broadening of the measured spectrum. Therefore, the energy shift parameter $\Delta_{10}$ can be constrained by the measured broadening $\sigma$:
\begin{equation}
    |\Delta_{10}| \leq \kappa_{1} \sigma,
    \label{eq:MachatschekIneq}
\end{equation}
where the coefficient $\kappa_{1}$ depends on the column density and can be calculated numerically~\cite{Machatschek2021_1000132391}. A similar constraint to equation~\ref{eq:MachatschekIneq} can be obtained for the twice- and thrice-scattered electrons with the corresponding energy shifts $\Delta_{20}$ and $\Delta_{30}$. An effective description can be used, replacing $\Delta_{i0}$ with a single $\Delta$ parameter for all the scatterings.

For KNM3-SAP and KNM3-NAP,4,5 direct measurements of the N$_{2,3}$-32 broadening are available for the tritium column densities of \SI{2.0e21}{\per\meter\squared} and \SI{3.8e21}{\per\meter\squared}, respectively. The corresponding squared broadenings are $\sigma^2_\mathrm{KNM3-SAP} =$ \SI{0.89\pm0.15e-3}{\electronvolt\squared} and $\sigma^2_\mathrm{KNM3-NAP,4,5} =$ \SI{0.89\pm0.19e-3}{\electronvolt\squared}. For KNM1 and KNM2 no krypton scans in the respective source configurations are available. The squared broadenings for these campaigns are obtained by extrapolation over the column density, $\sigma^2_\mathrm{KNM1} =$ \SI{0.90\pm0.90e-3}{\electronvolt\squared} and $\sigma^2_\mathrm{KNM2} =$ \SI{0.88\pm0.88e-3}{\electronvolt\squared}, and provide corresponding constraints for the $\Delta_{i0}$ parameters. The $\Delta_{i0}$ parameters are considered as fully correlated between the campaigns.

To ensure a stable source potential, a measurement procedure called an IU scan is performed on a daily basis. In an IU scan, the rear-wall bias voltage is ramped with a triangular modulation around the default set point. During this ramping, the currents at the rear wall and at the dipole electrodes inside the differential pumping section are recorded. For each voltage the sum of the electron and ion currents is obtained. The voltage with the total current of \SI{0}{\ampere} defines a so-called optimal rear-wall bias voltage without current at the walls of the source beam tube. The evolution of this optimal bias voltage provides information on the evolution of the source potential~\cite{Klein2019_1000093526,Friedel2020_1000126837}.

In addition to spatial and short-term time variations of the source potential, long-term changes are taken into account. They are obtained as the relative evolution of the effective endpoint of each scan, with the fits performed on the data excluding the neutrino-mass-sensitive analysis interval. The measured variance $\sigma^2_\mathrm{meas}$ of the fit parameter $E_0$ is compared to the expected variance $\sigma^2_\mathrm{expected}$. Any overdispersion is considered as additional broadening, $\sigma^2_\mathrm{drift}$, of the spectrum. For the KNM5 campaign this broadening is $\sigma^2_\mathrm{KNM5,drift} =$ \SI{8.6\pm0.9e-3}{\electronvolt\squared}. The values for the other campaigns are summarized in section~\ref{SubSec:ReleaseInputs}.

\subsection{Source magnetic field}
\label{SubSec:Bsource}

The magnetic field in the source, $B_\mathrm{src}$, enters the calculation of the response function, equation~\ref{Eq:ResponseFunction}, through the maximum acceptance angle, $\theta_\mathrm{max} = \arcsin{\sqrt{{B_\mathrm{src}}/{B_\mathrm{max}}}}$, and the transmission condition, $\mathcal{T}(E,qU,\theta)$. Therefore, $B_\mathrm{src}$ changes the shape of the modeled spectrum and can bias the neutrino-mass result. In the previous analyses~\cite{KATRIN:2019yun,KATRIN:2021uub} $B_\mathrm{src}$ was obtained by measurements of the stray field of the source superconducting solenoids with a Hall probe. An uncertainty of \SI{1.7}{\percent} was estimated by comparing the results to simulations.

\begin{figure}[!t]
    \centering
    \includegraphics[width=0.8\textwidth]{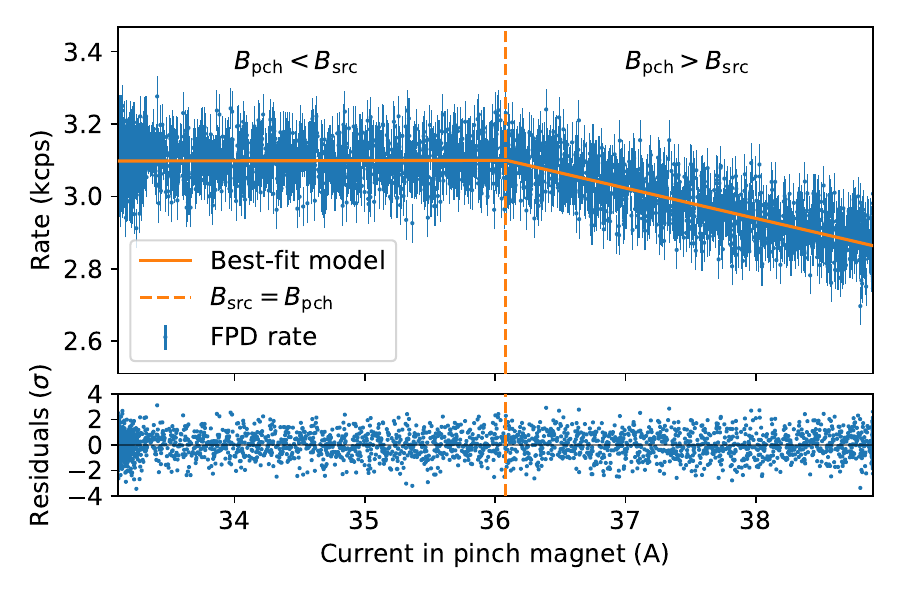}
    \caption{Measurement of the magnetic field in the source, $B_\mathrm{src}$. The measurement exploits the magnetic mirror effect and excellent knowledge of the maximal magnetic field in the beamline, created in the pinch magnet, $B_\mathrm{pch}$. The rate of electrons from the rear wall is measured at the FPD while the current in the pinch magnet is increased. As soon as the magnetic field in the pinch magnet is larger than the magnetic field in the source, the count rate of electrons starts to decrease, as electrons with a high pitch angle are reflected at the pinch magnet. The position of the dashed line is defined by a fit of the data with a piece-wise linear function. The lower panel shows the residuals of the fit, with a reduced $\chi^2$ of \num{0.99} for \num{2731} degrees of freedom.}
    \label{Fig:Bsrc}
\end{figure}

Recently, an improved measurement of the source magnetic field was performed~{\cite{Block2022_1000145073}}, based on the magnetic mirror effect and on the precise knowledge of the field in the pinch magnet, the superconducting solenoid at the exit of the main spectrometer that normally provides the highest magnetic field of \SI{4.2}{\tesla} in the beamline. The measurement is illustrated in figure~\ref{Fig:Bsrc}. The magnetic field in the source is fixed to its nominal value, and all the fields in the transport section are set to values below the source magnetic field of \SI{2.5}{\tesla} while the current in the pinch magnet is ramped up. The rate of \be-decay electrons from the rear wall is measured at a retarding energy of {\SI{16}{\kilo\electronvolt}}. The source is kept empty and the electrons from the rear wall probe the magnetic field. They are emitted isotropically in the magnetic field $B_\mathrm{RW} =$ \SI{1.23}{\tesla} and the highest magnetic field in the beamline defines the acceptance angle for these electrons. If the magnetic field in the pinch magnet is smaller than the source magnetic field, the maximum acceptance angle is $\theta_\mathrm{max} = \arcsin{\sqrt{\frac{B_\mathrm{RW}}{B_\mathrm{src}}}}$ and the same number of rear-wall electrons is transmitted. As soon as the $B_\mathrm{max}$ in the pinch magnet becomes larger than $B_\mathrm{src}$, the measured rate starts to decrease with the decreasing maximal acceptance angle. The position of the transition from the constant rate to a decreasing rate defines the current in the pinch magnet for which $B_\mathrm{src} = B_\mathrm{max}$.

With the new method described above, the maximal source magnetic field was determined to be $B_\mathrm{src} =$ \SI{2.513\pm0.003}{\tesla}. Comparing it to the simulated value of \SI{2.519}{\tesla} and adding the magnetic field uncertainty of the pinch magnet of \SI{0.1}{\percent}, the uncertainty of $B_\mathrm{src}$ was conservatively estimated at the level of \SI{0.25}{\percent}. 
The longitudinal variation of $B_\mathrm{src}$ along the \num{10}-m-long source leads to a slightly lower average magnetic field in the source. A simulation of the field profile in the source yields the value of $B_\mathrm{src}=\SI{2.507}{\tesla}$, which is used for all campaigns.

\subsection{Background-related systematic effects}
\label{SubSec:BackgroundSystematics}

\begin{figure}[!t]
    \centering
    \includegraphics[width=0.8\textwidth]{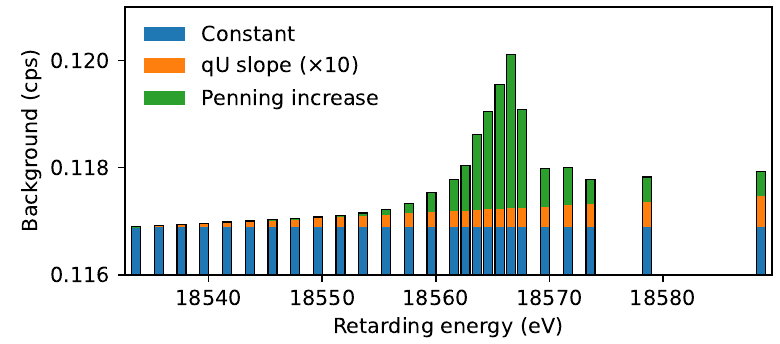}
    \caption{Background contributions. Most of the background is described by a constant rate. The additional components are dependent on the retarding energy and the scan-step duration, which are both constrained by dedicated analyses.}
    \label{Fig:Backgrounds}
\end{figure}

\begin{figure}[!t]
    \centering
    \includegraphics[width=0.8\textwidth]{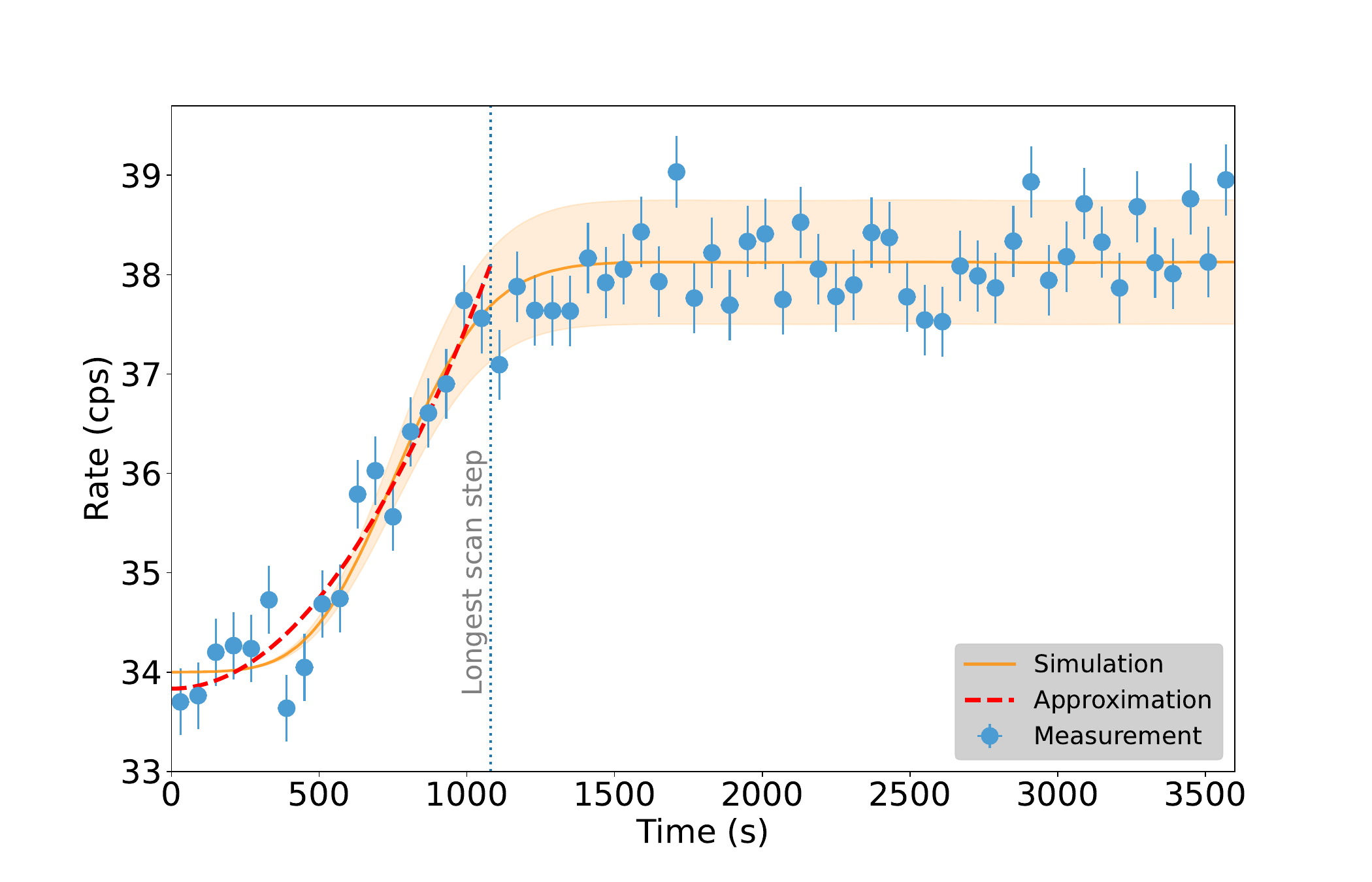}
    \caption{Rate evolution of the Penning-trap-related background. The measurement (blue points) is performed using a configuration with a low magnetic field of about $\SI{0.1}{\milli\tesla}$ in the main spectrometer. Due to the limited magnetic shielding, this configuration enhances the impact of the ions, which leave the trap and hit the spectrometer walls, producing background electrons. The solid orange line with the error band shows the overlay of the simulation of the Penning trap under realistic conditions. A quadratic function (dashed red curve) provides a robust approximation for all scan steps with a duration below \SI{1080}{\second}, using a minimal number of parameters.}
    \label{Fig:BackgroundsPenning}
\end{figure}

Figure~\ref{Fig:Backgrounds} shows the background contributions to the integrated tritium spectra. As most of the background does not depend on the retarding voltage of the spectrometers, it is primarily described by a constant rate. This rate is kept a free parameter for each data set, which accounts for changes in the surface conditions of the spectrometer since the last bake-out, the decay of \nuc{Po}[210], and spatial variations in the SAP configuration. It is largely constrained by the measurement points above the tritium endpoint.

A possible retarding-energy dependence, due to different transmission and detection probabilities of background electrons, is taken into account by an additional slope component. This component is constrained by dedicated calibration measurements of the spectrometer background in a wide range of retarding energies, $\SI{17.6}{\kilo\electronvolt}<qU <\SI{18.6}{\kilo\electronvolt}$, providing a slope of \SI{0.9\pm3.2}{\milli\cps\per\kilo\electronvolt} in the NAP setting and \SI{1.1\pm0.7}{\milli\cps\per\kilo\electronvolt} in the SAP configuration. The uncertainty is improved compared to previous data releases by additional calibration measurements.

The high voltage on both spectrometers generates an inter-spectrometer Penning trap, in which electrons accumulate and produce positive ions that can escape into the main spectrometer, where they can create low-energy background electrons. The trap is emptied between scan steps by a conductive wire~\cite{KATRIN:2019mkh}. This repeated reset leads to a scan-step-duration-dependent background that resembles the measurement-time distribution.  
Initially, the evolution of the additional rate from the Penning trap was described with a simple linear increase over time. However, simulations and dedicated measurements suggest a non-linear behavior of the rate with a sigmoid-like shape, see figure~\ref{Fig:BackgroundsPenning}. Since even for the longest scan step of \SI{1080}{\second} the plateau region is not reached, the evolution of the count rate of electrons due to the Penning trap is emulated by a quadratic function. Its coefficient is determined for each campaign and is taken into account using the time duration of each scan step. With equal scan steps in KNM4-OPT and the reduced potential of the pre-spectrometer, this effect is not present in KNM4-OPT and KNM5.

Radioactive decays in the spectrometer volume, predominantly due to \nuc{Rn}[219], produce \si{\kilo\electronvolt}-scale electrons that are trapped magnetically, lose energy in collisions with residual atoms, and generate correlated populations or clusters of secondaries. Typical clusters reach up to \num{100} events, arriving at the detector within up to \SI{1000}{\second}. These clusters generate a non-Poissonian overdispersion that is determined by an unbinned fit of a Gaussian distribution to the background data and comparing the variances of the Gaussian and Poisson fits. It yields an overdispersion $f_\mathrm{NP} =(\sigma_\mathrm{gaussian}^2/\sigma_\mathrm{Poisson}^2-1)$ of up to \num{0.1}. This is taken into account as an increased statistical uncertainty of the background rate. This effect is not observed for data taken in the SAP configuration, where the background-electron storage condition is altered.

\subsection{Residual tritium on the rear wall}
\label{SubSec:RWSpectrum}

\begin{figure}[!t]
    \centering
    \includegraphics[width=0.8\textwidth]{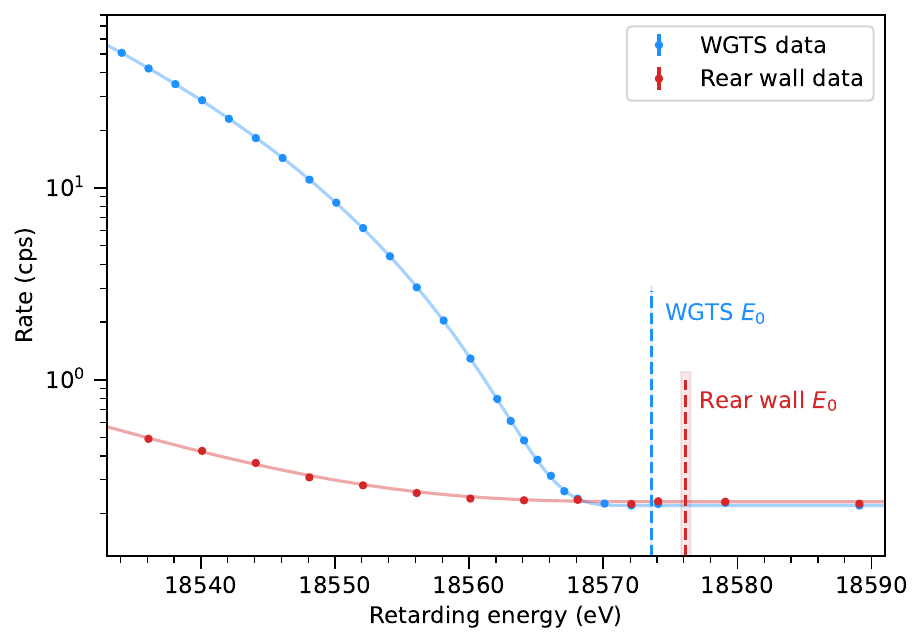}
    \caption{KNM3-NAP rear-wall electron spectrum in comparison to the WGTS spectrum. The signal amplitude is approximately \SI{1}{\percent} and the endpoint, $E_{0}$, is approximately \SI{2.5}{\electronvolt} higher (the shadowed regions represent the $E_{0}$ uncertainties). The spectrum of the tritium $\upbeta$-decay at the rear wall (red) is scanned in the same configuration as the WGTS spectrum, but with the evacuated/empty source. The WGTS spectrum (blue) combines the electrons from the rear wall and the $\upbeta$-decay electrons from the source after they pass through the (non-empty) source.}
    \label{Fig:RwSpectrum}
\end{figure}

As tritium is pumped through the KATRIN source, residual tritium is deposited onto the rear wall, dominantly through positive tritium ions coming from the source plasma. This residual tritium accumulates over time and leads to an underlying background tritium spectrum in neutrino-mass data. Scan measurements of the rear-wall tritium spectrum, where the source is empty with no circulating tritium, are performed intermittently throughout neutrino-mass measurement campaigns. One such measurement is shown in comparison to the KNM3-NAP spectrum in figure~\ref{Fig:RwSpectrum}. The signal amplitude of the rear-wall spectrum is approximately \SI{1}{\percent} that of the WGTS spectrum, with an endpoint approximately \SI{2.5}{\electronvolt} higher.

The rear-wall data sets are fit to obtain three spectral input parameters: the rear-wall endpoint, the rear-wall FSD shape, and the rear-wall signal amplitude. The endpoint of the residual tritium spectrum is expected to be higher than that of the gaseous molecular tritium in the WGTS because of the formation of carbon chains, most likely hydrocarbons or amorphous carbon from contaminants on the surface of the rear wall. Since the composition of contaminants is not known, and therefore the FSD is not precisely known, an additional parameter is included in the fit allowing additional freedom of the FSD shape. This shape parameter shifts the transition probability between the ground and excited states, and has a correlation coefficient with the endpoint of approximately \num{-0.96}, leading to a small increase in the systematic uncertainty on the neutrino mass.

\begin{figure}[!t]
    \centering
    \includegraphics[width=1.0\textwidth]{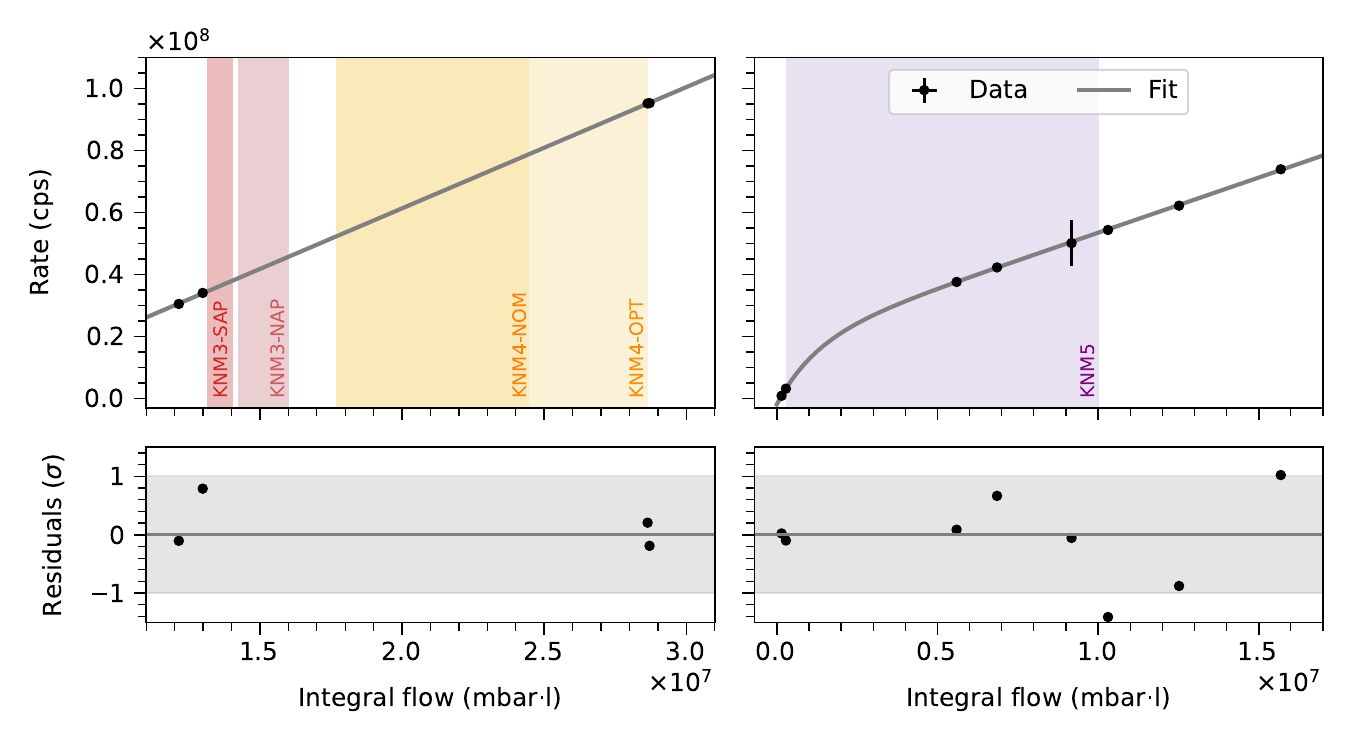}
    \caption{Measurements of the rear-wall activity evolution. The data is fit using a linear model before the rear-wall cleaning (left, KNM3,4) and with a limited-growth model after the cleaning (right, KNM5). The residuals of the fits are shown in the bottom panels. }
    \label{Fig:RwActivity}
\end{figure}

These spectral parameters are scaled based on the experimental conditions during the neutrino-mass measurement, and are subsequently used to model the rear-wall spectrum background in the neutrino-mass analysis. Since the signal amplitude increases over time due to the build-up of residual tritium, it is scaled using data from dedicated activity measurements to the time-weighted average of activity for each measurement campaign in order to reflect the rear-wall-spectrum background contribution during neutrino-mass measurements. A cleaning of the rear wall was performed using ozone cleaning between the KNM4 and KNM5 measurement campaigns, reducing the residual tritium activity on the rear wall by several orders of magnitude~\cite{Aker2023:OzonCleaning}. The build-up of residual tritium is expected to follow a limited-growth model, initially with a limited exponential increase of residual tritium on the surface and then converting to a linear increase. The signal-amplitude scaling is therefore done with a linear function before the rear-wall cleaning (where activity measurements were taken only in the linear part), and with a limited-growth model after the cleaning. The activity measurements and fits to the data used for the signal-amplitude scaling are shown in figure~\ref{Fig:RwActivity}.

\subsection{Detector-related effects}
\label{SubSec:Detector}

\begin{figure}[!t]
    \centering
    \includegraphics[width=1.0\textwidth]{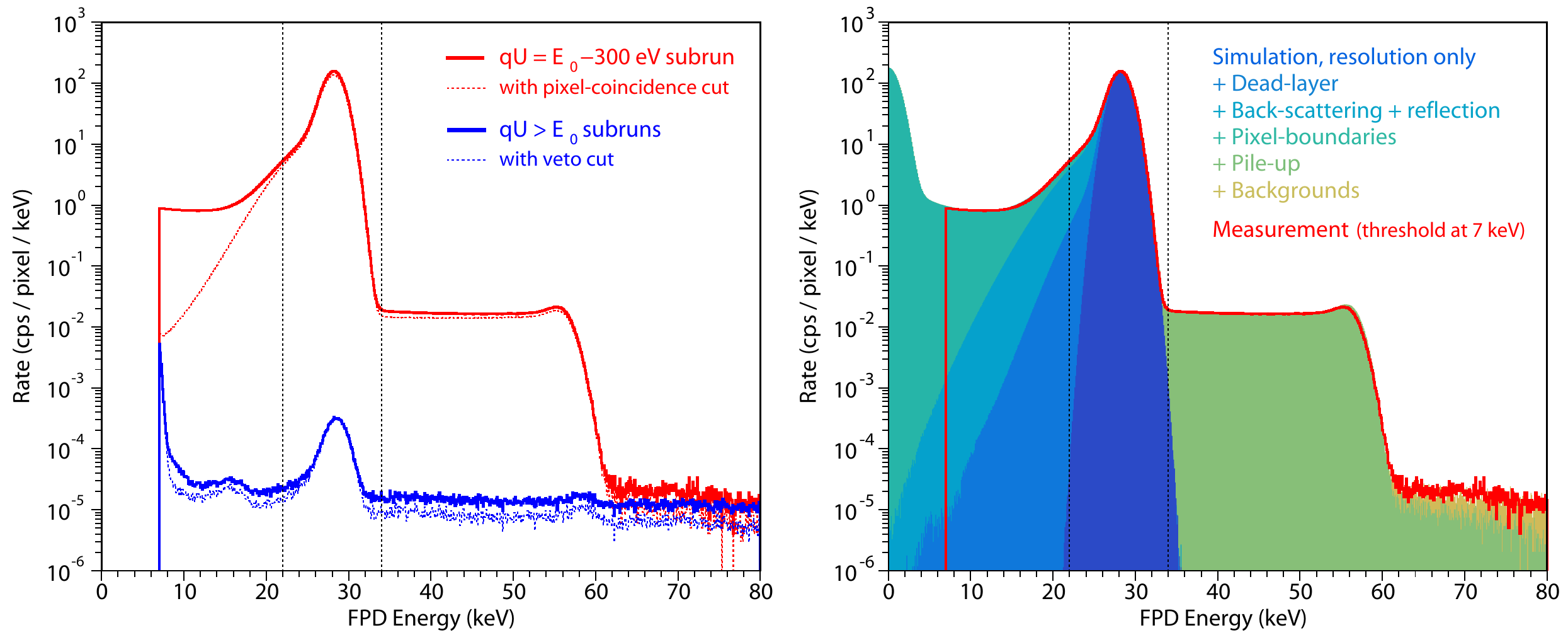}
    \caption{Left panel: the measured differential energy spectrum of electrons at the focal plane detector. The raw spectrum of the low-rate background HV set points (solid blue line) is compared to the same spectrum after application of the muon veto cut (dotted blue line), which further reduces the number of background events. The high-rate spectrum for the monitoring set point $qU = E_0 - \SI{300}{\electronvolt}$ is shown in red solid line for the raw measured rate and in red dashed line for the rate after applying the removal of coincidence events in different pixels. Right panel: Comparison of the simulation of the differential energy spectrum of the detector (shaded areas) with the measured spectrum (solid red line). The main effects are included in the simulation and added one by one to the modeled spectrum.}
    \label{Fig:Detector}
\end{figure}

Taking into account the post-acceleration energy of \SI{10}{\kilo\electronvolt}, electrons passing the main spectrometer hit the FPD pixels with about \SI{28}{\kilo\electronvolt} of energy, where the energy resolution is about \SI{3}{\kilo\electronvolt} (full-width at half maximum). Figure~\ref{Fig:Detector} displays a differential spectrum recorded at \SI{300}{\electronvolt} below the tritium endpoint, compared to a spectrum for background scans. This figure highlights the detector effects that govern the observed shape. The rate information for the integrated spectra is obtained by counting the events in a fixed region of interest (ROI) of \SIrange{22}{34}{\kilo\electronvolt} since KNM3, as opposed to \SIrange{14}{32}{\kilo\electronvolt} used in the first two campaigns. The change of the ROI improves the signal-to-noise performance. In addition, muon events and noise bursts are rejected. The detection efficiency $\epsilon_\mathrm{eff}$ after the ROI cut is about \SI{95}{\percent}. Its precise value is absorbed in the free normalization of the integrated spectrum fit. Only retarding-potential-dependent effects that change the detection efficiency are relevant.

With a decreasing retarding potential, lower-energy electrons pass the spectrometer and add counts to the detector differential spectrum, see figure~\ref{Fig:Detector}. The resulting change in the ROI coverage is modeled by an effective shift of the differential energy spectrum of the detector and is taken into account in the analysis. It amounts to an $\epsilon_\mathrm{eff}$ change of \SI{0.700\pm0.035}{\percent\per \kilo\electronvolt}. The uncertainty of this procedure is estimated from pixel-to-pixel variations of the differential spectrum. The impact on the squared neutrino mass is smaller than \SI{e-3}{\electronvolt\squared}, which is negligible for this analysis.

For lower retarding potential the rate and accordingly the probability for pile-up increase. As pile-up events likely appear outside the ROI, the detection efficiency is changed. This effect is taken into account using a rate-based model that includes the electronics response and signal processing. The efficiency correction at \SI{300}{\electronvolt} below the endpoint is  $\delta\epsilon_\mathrm{eff}=\SI{0.200\pm0.004}{\percent}$, which introduces an uncertainty of smaller than \SI{e-4}{\electronvolt\squared} for this analysis.

A significant fraction of the electrons that hit the detector undergo backscattering. They are then reflected by the electric and magnetic fields and hit the same or a different pixel within the shaping time. 
However, the lower the retarding potential the more electrons with larger surplus energies are present. Such electrons are more likely to overcome the spectrometer retarding potential in the direction of the source after backscattering at the detector.
Based on simulations, this backscattering loss of electrons is modeled as a \SI{0.10\pm0.02}{\percent\per\kilo\electronvolt} modification of the probability over the retarding energy. This estimate agrees well with the dedicated electron-gun measurements. The uncertainty is estimated by comparing the results from simulations performed with KESS~\cite{Renschler2011_1000024959} and GEANT4~\cite{GEANT4:2002zbu,Allison:2016lfl}. The corresponding impact on $m_\nu^2$ of less than \SI{e-3}{\electronvolt\squared} is negligible for this analysis. As the backscattering probability depends on the incident angle, and the repeated passage through the detector dead layer changes the measured spectral shape, a slight angular dependence is present. This effect is taken into account as a modification of the transmission function. The impact of this correction on the $m_\nu^2$ estimate is less than \SI{e-3}{\electronvolt\squared}.

Long-term changes in the gain are monitored in-situ using the \SI{300}{\electronvolt} monitoring point. However, changes within one scan may introduce a bias on the $m_\nu^2$. This potential bias has been estimated using an interpolation model between the scan-wise gain values and was found to be less than \SI{e-4}{\electronvolt\squared}.

%% file: 14sup-likelihood.tex
\section{Neutrino-mass inference}
\label{Sec:Supp-Likelihood}
This section describes the inference of the neutrino mass and other parameters from the KATRIN data.

\subsection{Parameter inference}
\label{SubSec:ParInference}

The squared neutrino mass $m^2_\upnu$ is inferred together with the other free parameters of the model ($E_0$, $A$, $R_\mathrm{bg}$) by simultaneously fitting the spectrum models of each individual campaign (equation~\ref{Eq:TotalSpectrum}) to the corresponding data using the method of maximum likelihood. The combined likelihood is constructed as a product of likelihood functions of each individual campaign: $\mathcal{L} = \prod_p \mathcal{L}_p$. The best-fit parameters are obtained by minimization of the negative logarithm of the likelihood, $-2\log \mathcal{L}=-2\sum_p \log \mathcal{L}_p$. Negative $\mnu^2$ values are allowed in the differential spectrum model in equation~\ref{Eq:DiffSpecT2}, with $\mnu=0$ in the Heaviside function's argument.

For the campaigns KNM1, KNM2, and KNM3-NAP, the Poisson distribution of measured counts at each data point with measurement time $t_\mathrm{meas}$ is approximated by a normal distribution with a mean of $\mu = R_\mathrm{calc}\cdot t_\mathrm{meas}$ and a variance of ${\mu}$. This approximation is valid for the campaigns performed in the normal-analyzing-plane configuration, for which the counts of all selected pixels are summed up and the number of counts for each data point is $\gtrsim10000$. This approximation also allows us to include the overdispersion of the background rate. The factor $f_\mathrm{NP}$ modifies the background variance at each measurement point $qU_i$, using $\sigma^2_{R_\mathrm{bg},i}=R_{\mathrm{bg},i}\cdot(1+f_\mathrm{NP})^2$. The resulting contribution to the minimized quantity is
\begin{equation}
    -2 \log \mathcal{L}_{\text{normal}} = \sum_{i}\frac{\left(R_{\mathrm{calc}}\left(qU_i\right) - R_{\mathrm{data}}\left(qU_i\right)\right)^2}{\sigma^2_{R,i}}.
    \label{Eq:GaussLikelihood}
\end{equation}

In the campaigns performed in the shifted-analyzing-plane configuration (KNM3-SAP, KNM4, and KNM5), the counts are split between \num{14} patches and the approximation of the Poisson distribution of counts with a normal distribution is no longer valid. Therefore, the Poisson likelihood is used, leading to the following contribution to $-2\log \mathcal{L}$:

\begin{equation}
    -2 \log \mathcal{L}_{\text{shifted}} = \sum_{i,k} 2 \left( R_{\mathrm{calc},k}(qU_i)\cdot t_i - N_{i,k} + N_{i,k}\cdot \ln{\frac{N_{i,k}}{R_{\mathrm{calc},k}(qU_i)\cdot t_i}} \right)
    \label{Eq:PoissonLikelihood},
\end{equation}
where the summation is performed over each of the \num{14} patches, $k = 0, 1.., 13$, and all HV set points $qU_i$.

The combined negative logarithm of the likelihood is given by
\begin{equation}
\begin{aligned}
    -2\log \mathcal{L_\mathrm{combined}}=& \sum_\mathrm{KNM1,2,3-NAP} \sum_{i}\frac{\left(R_{\mathrm{calc}}\left(qU_i\right) - R_{\mathrm{data}}\left(qU_i\right)\right)^2}{\sigma^2_{R,i}} +\\ 
    &\sum_\mathrm{KNM3-SAP,4,5} \sum_{i,k} 2 \left( R_{\mathrm{calc},k}(qU_i)\cdot t_i - N_{i,k} + N_{i,k}\cdot \ln{\frac{N_{i,k}}{R_{\mathrm{calc},k}(qU_i)\cdot t_i}} \right).
    \label{Eq:CombinedLikelihood}
\end{aligned}
\end{equation}
The total number of data points in the KNM1-5 campaigns in the analysis interval is \num{1609}. Therefore, each calculation of the likelihood in equation~\ref{Eq:CombinedLikelihood} requires \num{1609} evaluations of the model $R_\mathrm{calc}$. The evaluation of the model contains a nested integration, a summation over the final-states distribution, and the numerical solution of an equation, see section~\ref{Sec:Supp-TheoModelAndInputs}. A simple minimization typically needs thousands of $-2\log \mathcal{L}$ evaluations, while the estimation of uncertainties and evaluating the likelihood profiles becomes even more computationally expensive depending on the number of varied parameters in the fit.

To facilitate the analysis with respect to these computational requirements, two approaches were implemented by two independent analysis teams. In the first approach, the calculation of the likelihood in equation~\ref{Eq:CombinedLikelihood} is optimized with caching of multiple intermediate computation results, namely the scattering probabilities and the response function. The caching avoids re-calculation of all the components of the spectrum model at each step of minimization. This approach reduces the minimization time by about a factor of $\mathcal{O}(10^3)$.

The second approach exploits a fast prediction of the model with a neural network~\cite{Karl:2022jda} with spectrum model parameters ($E_0$, $m^2_\upnu$, column density, etc.) as the inputs and rates $R_\mathrm{calc}(qU_i)$ as the outputs. The network is trained on a large pre-calculated sample of spectrum-model values for various input parameters. After the training, the neural network can replace the model calculation with a precise prediction of the model over a wide range of input parameters. The evaluation of the neural network for each set of input parameters is also about a factor of $\mathcal{O}(10^3)$ faster than the usual code for model calculation. For each campaign (and each patch in SAP campaigns) a separate neural network is trained to ensure high accuracy of the model prediction. The networks are then used for the estimation of equation~\ref{Eq:CombinedLikelihood} and for inferring the fit parameters, including the squared neutrino mass.

\subsection{Systematic-uncertainty propagation}
\label{SubSec:SysPropagation}

The theoretical model of KATRIN's measured spectrum depends, apart from the free parameters $E_0, m^2_\upnu, A, R_\mathrm{bg}$, on multiple inputs that enter the experimental response or differential-spectrum calculation, see sections~\ref{Sec:Supp-TheoModelAndInputs} and \ref{Sec:Supp-Calibration}. The uncertainties of these additional nuisance parameters are included in the uncertainty estimation of the parameter of interest, $m^2_\upnu$. In the KATRIN analysis, the systematic uncertainty propagation is performed by the pull-term method.

In the pull-term method, all the additional nuisance parameters $\vec{\eta}_\mathrm{sys}$ of the model, for example, the column density $\rho d$, are considered as free fit parameters. They are, however, constrained by external measurements or simulations. Typically, the external information is given as a symmetric \SI{68.3}{\percent} CL confidence interval for each parameter, $\eta_\mathrm{sys,i} = \eta_\mathrm{ext,i}\pm \sigma_{\eta_\mathrm{ext,i}}$, or as a vector of $\vec{\eta}_\mathrm{ext}$ with a covariance matrix $\Theta_\mathrm{cov}$. A normal distribution, or a multivariate normal distribution, for these parameter estimators is assumed. The constraints are then included in the likelihood in equation~\ref{Eq:CombinedLikelihood} as additional terms
\begin{align}
    -2\log \mathcal{L}_\mathrm{sys,i} = & \frac{(\eta_{i}-\eta_\mathrm{ext,i})^2}{\sigma^2_{\eta_\mathrm{ext,i}}}; \\
    -2\log \mathcal{L}_\mathrm{sys} = & 
    \left(\vec{\eta}-\vec{\eta}_\mathrm{ext} \right)^T \cdot \Theta^{-1}_\mathrm{cov} \cdot \left(\vec{\eta}-\vec{\eta}_\mathrm{ext} \right).
    \label{Eq:PullTerms}
\end{align}
In the pull-term method, the systematic parameter estimation is informed by the external measurement as well as by the fit data. This results in a robust unbiased estimation of the parameters of interest. In total, the presented KATRIN analysis considers 144 constrained systematic parameters. The impact of each individual systematic effect to $m_\nu^2$ is obtained by comparing the uncertainty of the squared neutrino mass for the two cases: 1) the corresponding systematic parameters are free constrained parameters of the fit, and 2) the corresponding parameters are fixed to $\eta_\mathrm{ext,i}$.

The treatment of the overdispersion of the background rate is different due to the statistical nature of this effect. For the KNM1, KNM2, and KNM3-NAP campaigns the overdispersion is included as an increase of statistical uncertainty. The impact of this effect is estimated by performing the analysis with and without the overdispersion and comparing the uncertainties of the squared-neutrino-mass parameter.

\subsection{Likelihood profiles and uncertainty estimation}
\label{SubSec:LikelihoodProfiles}

To estimate the uncertainty of the squared neutrino mass, the negative logarithm of likelihood (equations~\ref{Eq:CombinedLikelihood} and \ref{Eq:PullTerms}) is profiled in the vicinity of its minimum. The likelihood profile is obtained by fixing $m_{\upnu}^{2}$ to a range of values and minimizing over all other parameters in each fit. The resulting $-2\log\left(\mathcal{L}\right)$ from each fit is then plotted over the range of fixed $m_{\upnu}^{2}$ values yielding the profile. For the individual measurement campaigns the $-2\log\left(\mathcal{L}\right)$ values are shown relative to the minimum, $\Delta \left( -2\log\left(\mathcal{L}\right)\right) = -2\log\left(\mathcal{L}\right) - \left( -2\log\left(\mathcal{L}\right)\right)_\mathrm{min}$.

\begin{figure}[!t]
    \centering
    \includegraphics[width=0.8\textwidth]{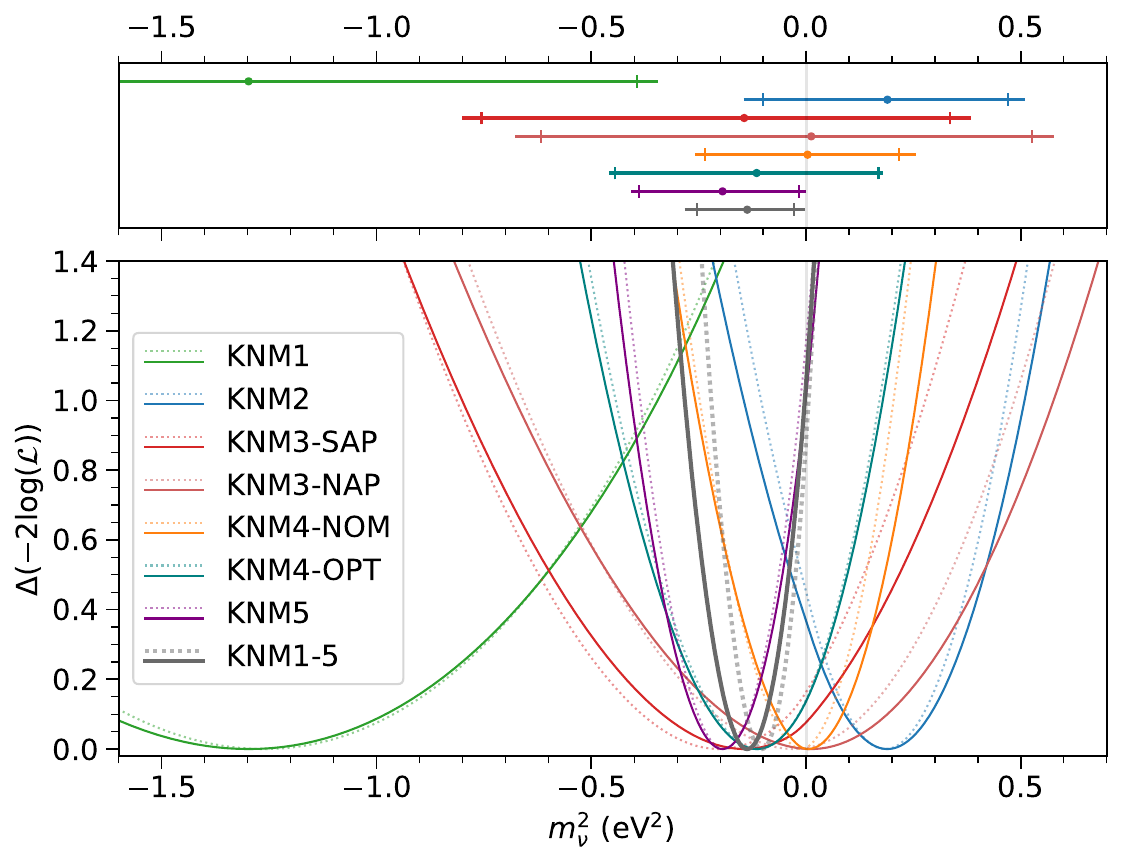}
    \caption{Likelihood profiles of each individual measurement campaign (colored) and the combined KNM1-5 analysis (gray). Statistics-only profiles are shown by the dotted lines; statistics and systematics profiles are shown by the solid lines. The top panel illustrates the central values of $m^2_\upnu$  with the \SI{68.3}{\percent} CL confidence intervals; the vertical bars indicate the statistics-only intervals.}
    \label{Fig:LikelihoodProfiles}
\end{figure}

The likelihood profiles for each individual campaign, as well as for the combined analysis, are shown in figure~\ref{Fig:LikelihoodProfiles}. The minimum of the fit including all systematics (solid lines) is very similar to the statistics-only minimum (dotted lines), and the profiles are widened only slightly since all campaign uncertainties are dominated by statistics. The top panel illustrates the central value of $m^2_\upnu$ for each campaign and for their combination, together with the $\SI{1}{\sigma}$ error bars (or \SI{68.3}{\percent} CL confidence intervals, which according to Wilks' theorem correspond to $\Delta \left( -2\log\left(\mathcal{L}\right)\right)=1$). 

\subsection{Confidence-interval construction}
\label{SubSec:ConfidenceInt}

A key result of the neutrino-mass measurement is the confidence interval for the parameter of interest, $m_\upnu$. The confidence intervals are constructed following a purely frequentist procedure by generating Asimov datasets with various true squared neutrino-mass values $m_{\upnu,\mathrm{true}}^{2}$, and fitting over a range of possible fit values $m_{\upnu,\mathrm{fit}}^{2}$ to obtain the likelihood profiles. All the constrained systematic parameters are included in the likelihood as described in section~\ref{SubSec:SysPropagation}. In the Feldman-Cousins construction the likelihood entries are ordered by likelihood ratio and summed until \SI{90}{\percent} coverage is reached. The left and right bounds of the acceptance regions are connected to construct the confidence belt~\cite{FeldmanCousins:1998}. The Feldman-Cousins approach provides a confidence interval in the physically allowed region of a bounded parameter. However, this prescription leads to stricter upper limits for unphysical negative squared neutrino mass values. 

To avoid such stricter limits, in the Lokhov-Tkachov approach, a symmetric acceptance region for $m_{\upnu,\mathrm{true}}^{2}$ values above the sensitivity is combined with a one-sided acceptance region for lower $m_{\upnu,\mathrm{true}}^{2}$~\cite{LokhovTkachov:2015}. In this construction, all negative (unphysical) $m_{\upnu,\mathrm{fit}}^{2} \le$ \SI{0}{\electronvolt\squared} yield the same confidence interval where the upper limit is the sensitivity and the lower limit is \SI{0}{\electronvolt\squared}. For the positive $m_{\upnu,\mathrm{fit}}^{2}$, the confidence interval has the same upper limit as the Feldman-Cousins method, while the lower bound has either the same or higher values.

\begin{figure}[!t]
    \centering
    \includegraphics[width=0.7\textwidth]{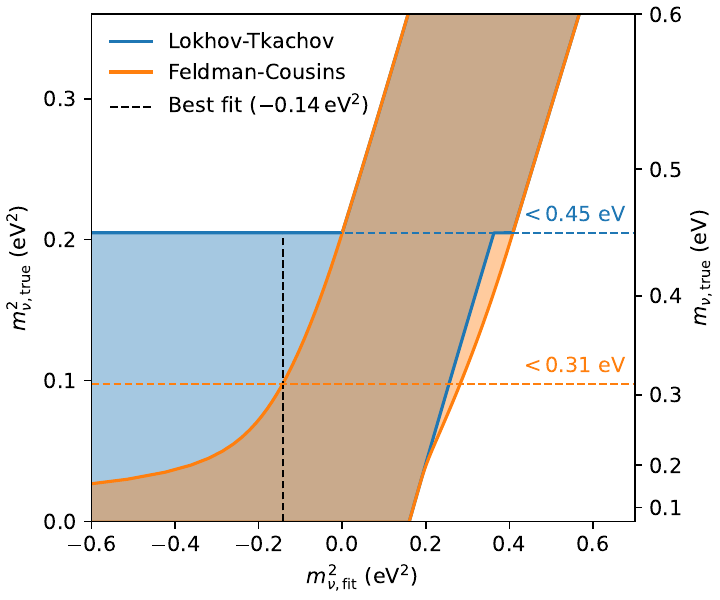}
    \caption{The construction of the confidence interval. The Lokhov-Tkachov construction is shaded in blue (solid blue boundaries), and the Feldman-Cousins construction is shaded in orange (solid orange boundaries) for comparison. The new upper limit on the neutrino mass from this work is $m_{\upnu} <$ \SI{0.45}{\electronvolt} (\SI{90}{\percent} CL).}
    \label{Fig:ConfidenceBelts}
\end{figure}

The two approaches to the confidence-belt construction for our present result are shown in figure~\ref{Fig:ConfidenceBelts}. To avoid the effect of obtaining shrinking upper limits when the squared neutrino-mass result statistically fluctuates into the unphysical negative region, we use the Lokhov-Tkachov construction as the default method. The corresponding neutrino-mass upper limit is $\mnu <$ \SI{0.45}{\electronvolt} at the \SI{90}{\percent} confidence level. This measurement represents the most stringent upper limit on the neutrino mass from a direct laboratory measurement.  For completeness, the Feldman-Cousins construction results in an upper limit of $\mnu <$ \SI{0.31}{\electronvolt} at the \SI{90}{\percent} confidence level, due to the slightly negative $m^2_\upnu$ best-fit value (illustrated by the dashed vertical line).

%% file: 15sup-campaigns.tex
\section{Individual campaigns and combination}
\label{Sec:Supp-IndividualCampaigns}

In this section, the main features of the individual and combined measurement campaigns are discussed. The fit results for the squared neutrino-mass parameter for each campaign and for their combination are shown in figure~\ref{Fig:LikelihoodProfiles}. The results for the other free parameters of the model are summarized in figure~\ref{Fig:FitParameters}.

\subsection{KNM1 and KNM2}
\label{SubSec:KNM12}

Multiple improvements are included in the present analysis with respect to \cite{KATRIN:2019yun,KATRIN:2021uub}. The extended calibration measurements described in section~\ref{SubSec:Plasma} allow us to reduce the uncertainty related to the source electric potential by a factor of \num{3.7}. The modified approach for measuring the column density yields a more robust estimation of this key systematic parameter, see section~\ref{SubSec:ColumnDensity}. A dedicated measurement of the retarding-energy-dependent background rate in the normal-analyzing-plane configuration, used in the KNM1 and KNM2 campaigns, results in a \SI{30}{\percent} smaller uncertainty contribution, see section~\ref{SubSec:BackgroundSystematics}.

The fits of the KNM1 and KNM2 campaigns with the updated inputs yield $\mnu^{2} = \SI[parse-numbers = false]{-1.30^{+0.95}_{-1.11}}{\electronvolt\squared}$ and $\mnu^{2} = \SI[parse-numbers = false]{0.19^{+0.33}_{-0.33}}{\electronvolt\squared}$, respectively. The best-fit values changed by less than \SI{20}{\percent} of the corresponding uncertainties compared to the previous results~\cite{KATRIN:2019yun,KATRIN:2021uub} due to the updated treatment of systematic effects and uncertainties.

\subsection{KNM3}
\label{SubSec:KNM3}

The KNM3 campaign was split into two measurement phases, during which two major modifications of the experimental configuration were applied. After the KNM2 campaign, the SAP configuration was shown to reduce the spectrometer background rate by a factor of two, and the method of in-situ measurement of the corresponding electromagnetic field was established, see section~\ref{Subsec:Supp-SAP}. The KNM3-SAP campaign demonstrated the feasibility of the tritium spectrum scans in the SAP configuration. To reduce possible impacts of the source electric potential, the column density was set to a lower value of $\rho d =$ \SI{2.05E21}{\per\meter\squared}, while the temperature of the source was set to \SI{80}{\kelvin}.

In the KNM3-NAP phase the column density of the source was increased to $\rho d =$ \SI{3.7E21}{\per\meter\squared} while keeping the temperature of the source at \SI{80}{\kelvin}. The higher temperature of the source compared to the KNM1 and KNM2 source temperature of \SI{30}{\kelvin} is used to ensure the same conditions in the source during \be-decay spectrum measurements and calibrations with the gaseous $^\mathrm{83m}$Kr admixture to tritium. The NAP spectrometer configuration was chosen to ensure the application of a well-established simpler data analysis, similar to the KNM1 and KNM2 campaigns.

Due to slight differences in the fit results of the two analysis teams for the KNM3-SAP and KNM3-NAP campaigns the best-fit values of each team are provided. The results for the KNM3-SAP campaign are $\mnu^{2} = \SI[parse-numbers = false]{-0.14^{+0.53}_{-0.66}}{\electronvolt\squared}$ and $\mnu^{2} = \SI[parse-numbers = false]{-0.16^{+0.55}_{-0.64}}{\electronvolt\squared}$. The fit of KNM3-NAP yields $\mnu^{2} = \SI[parse-numbers = false]{0.01^{+0.56}_{-0.69}}{\electronvolt\squared}$ and $\mnu^{2} = \SI[parse-numbers = false]{0.00^{+0.57}_{-0.69}}{\electronvolt\squared}$. The discrepancies are negligible with respect to the uncertainties.

\subsection{KNM4}
\label{SubSec:KNM4}

The KNM4 campaign combined the modifications of KNM3-SAP and KNM3-NAP: the spectrometer fields were set to the SAP configuration, while the source density was $\rho d =$ \SI{3.8E21}{\per\meter\squared} at a source temperature of \SI{80}{\kelvin}. This campaign was split into two periods: KNM4-NOM (for nominal measurement-time distribution) and KNM4-OPT (for the optimized time distribution).

During the data taking of KNM4-NOM, the duration of a single \be-spectrum scan was increased from \SI{2}{\hour} to \SI{3.5}{\hour}. Efficiencies of the scans were increased by reducing the relative amount of time spent on switching and stabilizing the HV set points. This updated measurement-time distribution contains points with measurement times above \SI{1000}{\second}, for which the increase of the Penning-trap-related background rate becomes significant. The additional background rate depends on the duration of the HV set point and can reach $\mathcal{O}(\SI{1}{\milli cps})$, see section~\ref{SubSec:BackgroundSystematics} and figure~\ref{Fig:Backgrounds}. To reduce the impact of this background, the measurement points with the longest duration were split into two shorter ones at the same HV set point at the end of KNM4-NOM. This reduces the impact of the time-dependent background rate on the measured spectrum.

A solution to mitigate the Penning-trap-related systematic effect was implemented in the later stage of the KNM4 campaign, KNM4-OPT. First, the measurement time at each point was split into equal intervals of \SI{100}{\second} and the Penning trap was emptied at the start of each interval using a special electron wiper~\cite{KATRIN:2019mkh}. In this way, the additional background rate from the Penning trap is the same for each HV set point, so it does not modify the shape of the measured spectrum and therefore does not affect the \mnusq estimate. In addition, the measurement time spent at different voltages was further optimized for better statistical sensitivity. During the last part of KNM4-OPT, the Penning trap was eliminated by decreasing the retarding potential of the pre-spectrometer from \SI{-10.5}{\kilo\volt} to \SI{-0.1}{\kilo\volt}. 

Modifications of the measurement times can introduce a bias in the rate estimation when combining data sets with changing operational parameters. During the first analysis of KNM4-NOM and KNM4-OPT, the drift of the source potential of about $\SI{60}{\milli\volt}$ was not taken into account and led to imperfect modeling of the spectral rate. The impact on the squared neutrino mass estimated by simulation is of the order of $\mathcal{O}(\SI{0.1}{\electronvolt^2})$, which is well below the statistical uncertainty of the KNM4 data set. Several post-unblinding tests triggered by the search for light sterile neutrinos have revealed this issue and resulted in a careful re-evaluation of all the systematic contributions and analysis procedures. As a result, the campaign was split into the two above-mentioned periods, each analyzed as an independent data set with a common \mnusq but different effective endpoint, background, and normalization parameters.
The best-fit of the squared neutrino mass for KNM4 is $\mnu^{2} = \SI[parse-numbers = false]{-0.05^{+0.19}_{-0.22}}{\electronvolt\squared}$. The best fits of KNM4-NOM and KNM4-OPT are $\mnu^{2} = \SI[parse-numbers = false]{0.01^{+0.25}_{-0.27}}{\electronvolt\squared}$ and $\mnu^{2} = \SI[parse-numbers = false]{-0.12^{+0.30}_{-0.34}}{\electronvolt\squared}$ respectively.

\subsection{KNM5}
\label{SubSec:KNM5}

By the end of the KNM4 campaign, the rear wall had been exposed to an integral flow of \SI{2.9e7}{\milli\bar \cdot l} of tritium; the total activity of tritium on the rear wall was above \SI{1}{\percent} of the tritium activity in the source. To minimize the impact of the corresponding systematic effect, the rear wall was de-contaminated before the start of the KNM5 campaign using ultraviolet irradiation and ozone cleaning~\cite{Aker2023:OzonCleaning}. The amount of tritium decays at the rear wall was reduced by three orders of magnitude.

The de-contamination of the rear wall changes its work function. Therefore, after restarting the tritium circulation, an initial drift of the work function of the rear-wall surface, and thus a drift of the source potential, was observed. As described in section~\ref{SubSec:Plasma}, the long-term evolution of the source potential is monitored via the IU scans. The bias voltage applied to the rear wall is used to correct for the significant changes of the source potential and minimize spatial variations. A drift of the starting potential of the order of $\mathcal{O}(\SI{100}{\milli\electronvolt})$ was observed at the beginning of the KNM5 campaign. It is taken into account by the additional broadening parameter $\sigma^2_\mathrm{KNM5,drift}$ as described in section~\ref{SubSec:Plasma}.

The best-fit of the squared neutrino mass for KNM5 is $\mnu^{2} = \SI[parse-numbers = false]{-0.19^{+0.19}_{-0.21}}{\electronvolt\squared}$.
Apart from the source-potential changes at the start of the campaign, KNM5 is the most stable among the first five data sets. 
In the subsequent measurement campaigns of KATRIN, the KNM5 configuration -- with the shifted analyzing plane and the high density source at \SI{80}{\kelvin} -- is used.

\subsection{Combined analysis}
\label{SubSec:CombinedAnalysis}

The combined fit of the KATRIN model to the data of first five campaigns is performed by minimizing the negative logarithm of the likelihood (equation~\ref{Eq:CombinedLikelihood}) including the systematic pull terms (equation~\ref{Eq:PullTerms}).
To take into account possible variations of the effective endpoint, signal normalization, and the background rate ($E_0, A, R_\mathrm{bg}$), these parameters are treated as independent free parameters for each campaign and for each detector patch in KNM3-SAP, KNM4-NOM, KNM4-OPT, and KNM5. Including the squared neutrino-mass parameter, there are $N_{\mathrm{free}} = 1 [m^2_\upnu] + 3 [E_{0,\mathrm{KNM1,2,3-NAP}}] + 3 [A_{\mathrm{KNM1,2,3-NAP}}] + 3 [R_{\mathrm{bg,KNM1,2,3-NAP}}] +\\ 4\cdot 14 [E_{0,\mathrm{KNM3-SAP,4-NOM,4-OPT,5}}] + 4\cdot 14 [A_{\mathrm{KNM3-SAP,4-NOM,4-OPT,5}}] +\\ 4\cdot 14 [R_{\mathrm{bg,KNM3-SAP,4-NOM,4-OPT,5}}] = 178$ free fit parameters.

\begin{figure}[!t]
     \centering
     \includegraphics[width=0.8\textheight,angle=90]{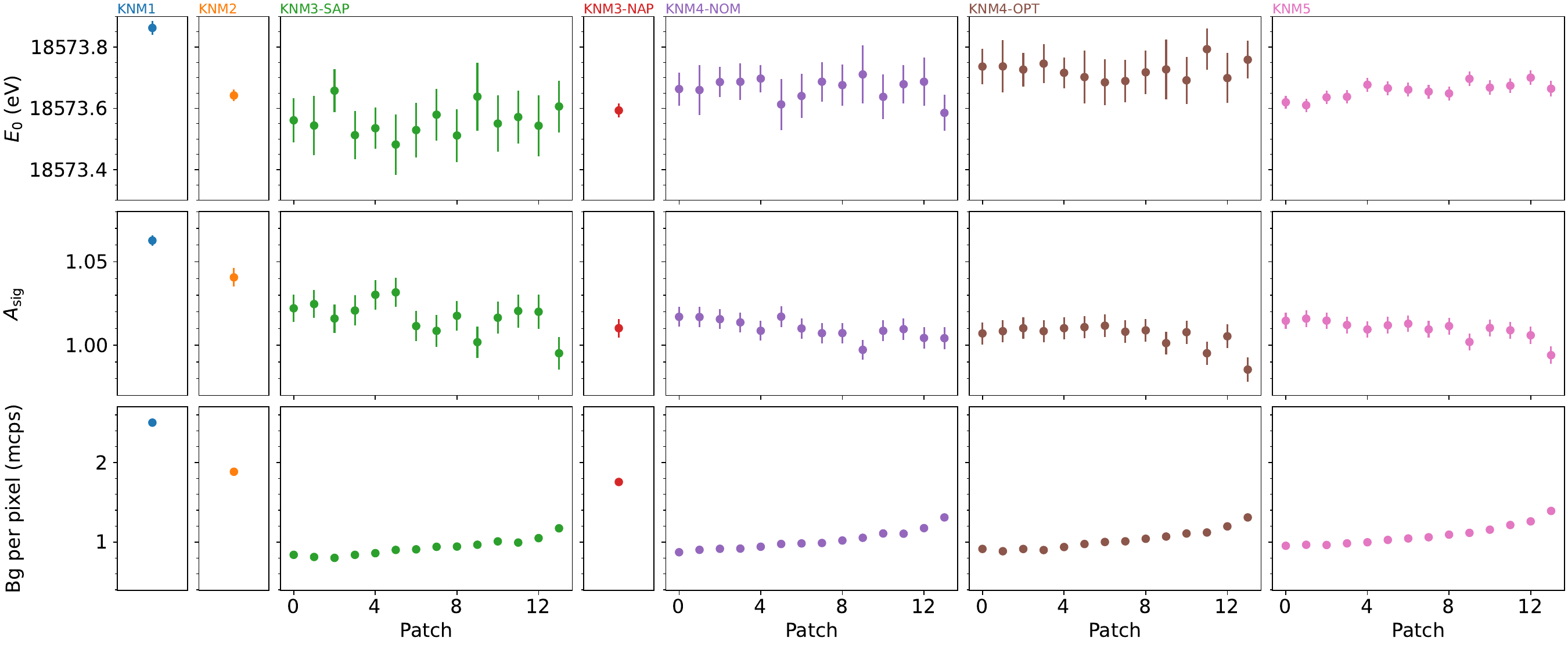}
     \caption{Fit results of the free fit parameters for each campaign. The best-fit values and \SI{1}{\sigma} uncertainties of the effective endpoint (top), signal normalization (middle), and the background rate (bottom) are shown. These fit results are for the active pixels in each measurement campaign.
     }
     \label{Fig:FitParameters}
 \end{figure}

\begin{figure}[!t]
     \centering
     \includegraphics[width=0.5\textwidth]{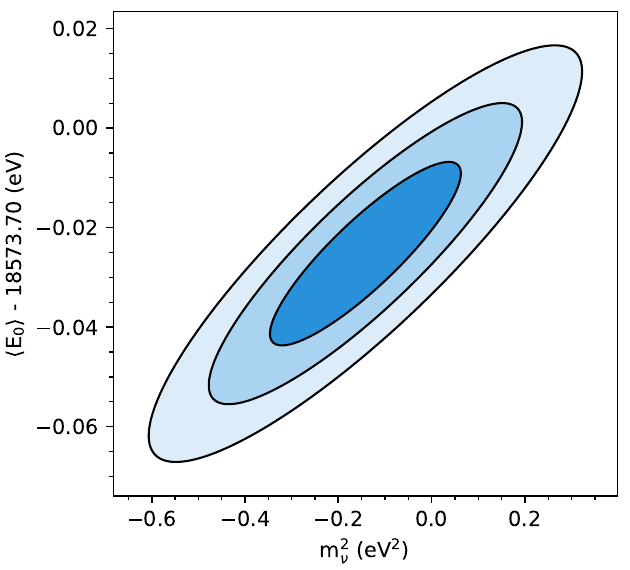}
     \caption{Correlation of $m_{\upnu}^{2}$ and $E_{0}$. The plot shows the 1, 2, and \SI{3}{\sigma} contours for the squared neutrino mass and the effective endpoint averaged over all the campaigns and \num{14} patches.}
     \label{Fig:Correlation}
 \end{figure}

The best-fit value for the squared neutrino mass \mbox{$\mnu^{2} = \SI[parse-numbers = false]{-0.14^{+0.13}_{-0.15}}{\electronvolt\squared}$} was obtained by the two independent analysis teams. The effective endpoint $E_0$ is correlated to $m^2_\upnu$ in the fit; the correlation is shown in figure~\ref{Fig:Correlation} where multiple endpoint values are combined as a correlated weighted mean. The correlations between $m^2_\upnu$ and $A$ or $R_\mathrm{bg}$ are significantly smaller. The fit values of the effective endpoint for each campaign and patch are shown in the top row of figure~\ref{Fig:FitParameters}. The variations of the effective $E_0$ values between the campaigns are related to the changes in the work functions of the source and the spectrometer and the corresponding evolution of the energy scale. The slight radial variation of $E_0$ for KNM5 is caused by the suboptimal rear-wall bias used in this campaign. The Q-value estimation is discussed in section~\ref{Sec:Supp-Endpoint}. 

The signal normalization parameter $A$ (middle row, figure~\ref{Fig:FitParameters}) shows a slight radial dependency, which could be attributed to the radially dependent absolute detection efficiency. The bottom row of figure~\ref{Fig:FitParameters} shows the background rate estimates for all the campaigns. The background is expected to increase for the patches in the outer part of the detector~\cite{KATRIN:2020zld}.

As described in section~\ref{SubSec:SysPropagation}, the systematic parameters, together with their uncertainties, are included in the fit. For the model inputs that are the same for each campaign (for example, the energy-loss function) their common parameters are used in the likelihood evaluation. Other inputs are obtained independently for each campaign and are treated as individual parameters during the fit. Another group of systematic inputs consists of the parameters, which are obtained from a separate measurement for each campaign but using the same method. The uncertainties inherent to this method lead to a partial correlation of these parameters between the different campaigns. The full list of input parameters with their uncertainties and correlations is given in section~\ref{Sec:Supp-DataRelease}. 

%% file: 16sup-endpoint.tex
\section{Q-value determination}
\label{Sec:Supp-Endpoint}

An estimation of the Q-value for the decay of molecular tritium provides an independent consistency check of the energy scale of KATRIN. The measured Q-value is compared to the value obtained from the Penning-trap measurement of the atomic-mass difference of tritium and helium-3, $Q(T_2)_{\Delta M} =$ \SI{18575.78\pm0.02}{\electronvolt}~\cite{MedinaRestrepo:2023qbj}, taking into account the ionization and molecular dissociation energies~\cite{Otten:2008zz}.

To obtain the Q-value from the \be-spectrum of KATRIN, one sums up the best-fit result for the effective endpoint $E_{0,\mathrm{obs}}$, corrected for the recoil of the $^3$HeT$^+$ molecular ion ($E_\mathrm{rec} =$ \SI{1.72}{\electronvolt},~\cite{Otten:2008zz}), the electron starting potential in the source, and the work-function difference between the source and the spectrometer. The last two contributions can be determined by comparing the line positions of $^\mathrm{83m}$Kr conversion electrons measured in KATRIN to the literature values. The difference between the measured L$_3$-32 line position $\mu_\mathrm{obs}$ and the literature value $\mu_\mathrm{ref} =$ \SI{30471.9(3)}{\electronvolt}~\cite{Slezak:2012zz,Inoyatov:2023vjj, McCutchan:2015vcl} is added to the observed endpoint value
\begin{equation}
    E_0 = E_{0,\mathrm{obs}} + \mu_\mathrm{ref} - \mu_\mathrm{obs}.
\end{equation}
The linearity of the energy scale and the high accuracy of the spectrometer high-voltage system~\cite{Rest_2019} allows to measure at the L$_3$-32 line and at the tritium endpoint energy.
A different approach was used for the KNM1 and KNM2 campaigns, based on the direct measurement of the work functions of the source, the rear wall and the spectrometer as described in~\cite{KATRIN:2021fgc}.

The Q-value estimations for all measurement campaigns are consistent with each other. The uncertainties are dominated by either the extrapolation of the work-function evolution over a long time (KNM1 and KNM2) or by the uncertainty of the literature reference value for the L$_3$-32 line position (KNM3, KNM4, and KNM5). The KNM4-NOM campaign was chosen to obtain the Q-value of KATRIN since the additional corrections to the Q-value analysis are the smallest, due to the source-potential stability at the level of \SI{30}{\milli\electronvolt}. The observed endpoint, $E_\mathrm{0,obs,KNM4-NOM} =$ \SI{18573.66\pm0.01}{\electronvolt}, is obtained as the mean of the best-fit effective endpoint values of all patches, see figure~\ref{Fig:FitParameters}. The measured line position after the corrections is $\mu_\mathrm{obs} =$ \SI{30472.25\pm0.05}{\electronvolt}. By combining $E_0,\mu_\mathrm{obs},\mu_\mathrm{ref}$, and $E_\mathrm{rec}$ one obtains a Q-value of $Q_\mathrm{KATRIN} =$ \SI{18575.0\pm0.3}{\electronvolt}, which is in slight (about $2.5 \sigma$) tension with the Penning-trap measurements. Further improvement of the Q-value determination is possible using the transition energies of $^\mathrm{83m}$Kr~\cite{Rodenbeck:2022fxc}.

%% file: 16_5sup-postunblinding.tex
\section{Post-unblinding modifications}
\label{Sec:Supp-PostUB}

To ensure the transparency of our analysis procedures, we list all the modifications of the analysis that were implemented after the initial unblinding of the data. For each modification, the impact on the neutrino mass is estimated.

\begin{itemize}
    \item Data combination: The KNM4 campaign was split into two periods, KNM4-NOM and KNM4-OPT, according to the distribution of scan-step durations. This division allows the proper accommodation of different effective endpoint values caused by a drift of the starting potential in the source. This modification causes a shift of the squared neutrino-mass value of the KNM4 campaign by about $\SI{-0.1}{\electronvolt^2}$. In the future, such effects of data combination will be addressed by additional analysis steps on the simulated data.
    \item Column density: With the re-evaluation of the key systematic parameters, triggered by the mistake in the data combination, and measurements with the new monoenergetic photoelectron source, an additional correction of the column density was introduced. It takes into account the dependency of the electron angles on its starting kinetic energy. Since this correction was found a posteriori, it is included into the column density estimation based on an estimate from a single measurement with the old setup. With no possibility to test the effect for the various old configurations of the electron-gun setup, the total size of the effect is considered as an additional independent uncertainty. The corresponding impact on the squared neutrino mass is about $\SI{-0.05}{\electronvolt^2}$.
    \item Energy-loss function: Calibration measurements with the new photoelectron source, installed at the rear section of the KATRIN setup in 2022, pointed to a possible discrepancy between the two measurement modes: integral and time-of-flight, described in~\cite{KATRIN:2021rqj}. The different energy-loss parameters obtained in these two modes may induce a bias of the squared neutrino mass of up to $\SI{0.035}{\electronvolt^2}$. The uncertainty of the energy-loss parameters was increased by scaling the corresponding covariance matrix elements by a factor of 80 to cover this potential bias. The origin of the discrepancy between the two measurement modes is under investigation. Since the additional systematic uncertainty on the neutrino mass squared of $\SI{0.035}{\electronvolt^2}$ has no significant impact on the current sensitivity, we use the simple covariance scaling for this analysis and aim to resolve this discrepancy for future analyses.
    \item Penning-trap-related background: The post-unblinding re-evaluation revealed strong hints for a non-linear dependence of the background rate on the scan-step duration. Hence, the model was changed from a linear function to a quadratic increase. The possible bias in comparison to a many-parameter sigmoid model was found to be negligible. The corresponding impact on \mnusq is about $\SI{0.05}{\electronvolt^2}$ and is smaller than the uncertainty associated with this systematic effect.
    \item Rear-wall residual tritium: While revising the systematic uncertainties several minor discrepancies in the rear-wall spectrum analysis were noticed and corrected. The impact of the corrections on the squared neutrino mass is below the level of \SI{0.001}{\electronvolt^2}.
\end{itemize}

The initial best-fit value of the squared neutrino mass was $\mnu^{2} = \SI[parse-numbers = false]{-0.05^{+0.11}_{-0.12}}{\electronvolt\squared}$. The corresponding upper limit of $\mnu < \SI{0.43}{\electronvolt}$ at \SI{90}{\percent} CL was obtained.
After including the above modifications, the final best-fit value is $\mnu^{2} = \SI[parse-numbers = false]{-0.14^{+0.13}_{-0.15}}{\electronvolt\squared}$ with the upper limit of $\mnu < \SI{0.45}{\electronvolt}$ at \SI{90}{\percent} CL.

The above-listed post-unblinding modifications were included as corrections and improvements of the initial analysis procedure and inputs. None of the post-unblinding decisions were taken based on the neutrino mass best-fit values, which leaves our main neutrino-mass estimation bias free.

%% file: 17sup-release.tex
\section{Data release}
\label{Sec:Supp-DataRelease}

This section summarizes the released data, separated into the spectral measurement data and the input parameters to construct the model."

\subsection{Spectral measurement}

The KATRIN data of the five measurement campaigns, which were used for the neutrino-mass analysis presented in this paper, can be found in the file \textbf{{KATRIN\_data\_KNM1\-5.json}}. As described in section~\ref{Sec:Supp-DataProcessingSelectionCombination}, all the \be-decay scans are stacked by summing up the counts and the measurement time, and averaging the retarding voltages for each HV set point. The counts for all the golden pixels in the KNM1, KNM2, and KNM3-NAP campaigns are also summed, resulting in one number of counts per HV set point. The counts for all pixels in each patch in KNM3-SAP, KNM4-NOM, KNM4-OPT, and KNM5 are summed, yielding \num{14} count tallies per HV set point.

For each set point the following information is provided:
\begin{itemize}
    \item \textit{Retarding\_voltage}: the voltage applied to the main spectrometer in volts;
    \item \textit{Live\_time}: total measurement time at the given HV set point in seconds;
    \item \textit{Penning\_duration}: the average duration of the measurement at the given HV set point in each run of the campaign (in seconds), see section~\ref{SubSec:BackgroundSystematics};
    \item \textit{Penning\_total\_duration}: the total time of the measurement at the given HV set point with a given \textit{Penning\_duration} (in seconds). It is used in KNM4-NOM, where the measurement time distribution was modified twice. The rate contribution from the Penning trap is a time-weighted sum of the contributions from each period of KNM4-NOM.
    \item \textit{Event\_counts}: the total number of events recorded by the detector;
    \item \textit{Relative\_efficiency}: the relative detection efficiency, see section~\ref{SubSec:Detector};
\end{itemize}

\subsection{Model and systematic inputs}
\label{SubSec:ReleaseInputs}

The systematic inputs of the KATRIN \be-spectrum model with their uncertainties and correlations are collected in a single file, \textbf{KATRIN\_inputs\_KNM1\-5.json}. In this section the structure of various input entries is described.

\noindent\textbf{Electromagnetic fields}. The following six parameters enter the calculation of the KATRIN spectrum model
\begin{itemize}
    \item Magnetic field in the source $B_\mathrm{src}$, entry \textit{MagneticFieldSource}, has the same value for all the campaigns. In a combined fit it should be treated as a single parameter with the corresponding uncertainty.
    \item Maximal magnetic field $B_\mathrm{max}$, entry \textit{MagneticFieldMax}. The values and uncertainties contain \num{17} entries: a single value for KNM1, KNM2, and KNM3-NAP (elements \textit{knm1}, \textit{knm2}, \textit{knm3b}) and \num{14} values for patches in KNM3-SAP, KNM4, and KNM5 (elements \textit{knm3a45-patch00}, etc.).
    \item Magnetic field in the analyzing plane $B_\mathrm{ana}$, entry \textit{MagneticFieldAna}. $B_\mathrm{ana}$ is defined as a single value for KNM1, KNM2, and KNM3-NAP, and \num{14} values for the patches of KNM3-SAP, KNM4 and KNM5.
    \item Squared broadening of the transmission $\sigma^2$, entry \textit{TransmissionBroadeningVariance}. These are single values for the variance of KNM1, KNM2, and KNM3-NAP, and \num{14} values for the patches of KNM3-SAP, KNM4, and KNM5. The uncertainties of the variance are set to \textit{null} for the KNM1, KNM2, and KNM3-NAP.
    \item The retarding potential in the analyzing plane is given by the entry \textit{RetardingVoltageOffset}. The provided values are the differences between the retarding energy of the spectrometer and the actual retarding energy in the analyzing plane. For the KNM3-SAP, KNM4, and KNM5 campaigns, the corresponding values are shown in figure~\ref{Fig:SAP} as potential depression. The uncertainty of this parameter is not provided since it only affects the estimation of the effective endpoints $E_0$, but not \mnusq.
\end{itemize}

The following correlations between the parameters are taken into account. The uncertainties of the $B_\mathrm{ana}$ of the KNM1, KNM2, and KNM3-NAP campaigns are obtained with the same comparison of the field measurements and simulation, and are therefore fully correlated, see the correlation matrix in the entry \textit{MagneticFieldAnaKNM123b}. The correlations of the $B_\mathrm{ana}$ and transmission broadening $\sigma^2$ are given in the two correlation matrices in the entries \textit{AnalyzingPlaneKNM3a4} and \textit{AnalyzingPlaneKNM5}, their correlations are shown in figure~\ref{Fig:SAP}. Only the uncertainties of the parameters within one patch are correlated; the other off-diagonal elements in the matrices are zero. Finally, the uncertainties of the $B_\mathrm{max}$ parameter are strongly correlated between KNM1, KNM2, and KNM3, KNM4, and KNM5 values (entry \textit{Correlations} $\rightarrow$ \textit{MagneticFieldMax}), because the main part of the uncertainty of $B_\mathrm{max}$ is shared between all the campaigns.

\noindent\textbf{Column density.} The column-density parameters are provided for each campaign in the entry \textit{GasDensity}. The uncertainties and correlations are given by the covariance matrix, see section~\ref{SubSec:ColumnDensity} for more details.

\noindent{\textbf{Tritium purity.}} The numbers of the T$_2$, HT, and DT molecules in the source
$N_\mathrm{T_2}$, $N_\mathrm{HT}$, and $N_\mathrm{DT}$ are provided in the entry \textit{Concentrations} for each campaign. They can be converted into two parameters: the tritium purity $\varepsilon_\mathrm{T}=\left[ N_\mathrm{T_2} + \frac{1}{2}\left( N_\mathrm{HT}+N_\mathrm{DT} \right) \right]/N_\mathrm{tot}$ and HT/DT ratio $\kappa = N_\mathrm{HT}/N_\mathrm{DT}.$
The tritium purity evolution in all campaigns is shown in figure~\ref{Fig:SlowControl}. Due to the stability of these parameters at the level of $\mathcal{O}(1)~\%$, the average concentrations and corresponding values of $\varepsilon_\mathrm{T}$, $\kappa$ are used for each campaign with a negligible impact on \mnusq.

\noindent\textbf{Parameters of the rear-wall \be-spectrum.} As described in section~\ref{SubSec:RWSpectrum} the spectrum of the electrons, emitted in the \be-decay of tritium at the rear wall, is described by three effective parameters: the endpoint $E_\mathrm{0,RW}$, the amplitude of the spectrum, and the ratio of the ground and electronic excited final-states probabilities. The endpoint (entry \textit{EndpointRearWall}) is defined individually for KNM3, KNM4, and KNM5. The ratio (\textit{FSDShapeRearWall}) is estimated for KNM3-SAP, KNM3-NAP, KNM4, and KNM5. The amplitude (\textit{AmplitudeRearWall}) is given for each patch of KNM3-SAP, KNM4-NOM, KNM4-OPT and KNM5, while a single value for all golden pixels is provided for the KNM3-NAP campaign.

The full correlations of the amplitude parameters for the same patches of KNM3a and KNM4 are presented in \textit{Correlations} $\rightarrow$ \textit{Amplitude} $\rightarrow$ \textit{CorrelationMatrix}. The correlations between the ratio and the endpoint parameters for different campaigns are given by \textit{Correlations} $\rightarrow$ \textit{EndpointShape} $\rightarrow$ \textit{CorrelationMatrix}.

\noindent\textbf{Background.} The background contribution to the KATRIN spectrum is defined by three main effects, see section~\ref{SubSec:BackgroundSystematics}. The scan-step-duration-dependent contribution from the inter-spectrometer Penning trap is provided for KNM1, KNM2, KNM3-SAP, KNM3-NAP, and KNM4-NOM (entry \textit{Penning}). The Penning-trap effect was removed in the KNM4-OPT and KNM5 campaigns by equal scan-step durations and by lowering the pre-spectrometer voltage. The background-rate dependency on the retarding energy (\textit{BackgroundSlope}) was measured independently for KNM1,2,3-NAP and KNM3-SAP,4,5 due to the different electromagnetic-field settings in these campaigns. Finally, for the campaigns performed in the symmetric spectrometer-field configuration, KNM1, KNM2, and KNM3-NAP, the estimated overdispersion of background rate fluctuations is given by entry (\textit{BackgroundOverdispersion}). The relative overdispersion (\textit{knm1-relative}) is provided together with the \textit{knm1-sigma}, \textit{knm1-meancounts}, \textit{knm1-meantimes} values which are required for estimating the overdispersion contribution for different durations of HV set points.

\noindent\textbf{Energy-loss function.} The parameterization of the energy-loss function and the input parameters originate from the energy-loss function measurement~\cite{KATRIN:2021rqj}. The same parameters (entry \textit{EnergyLoss} $\rightarrow$ \textit{Value}) are used for all the measurement campaigns.
The modified covariance matrix used in the analysis is provided in the entry \textit{EnergyLoss} $\rightarrow$ \textit{CovarianceMatrix}.

\noindent\textbf{Angular-dependent detection efficiency.} The phenomenological description of the angular-dependent detection efficiency, see sections~\ref{Sec:Supp-TheoModelAndInputs} and \ref{SubSec:Detector}, is used to correct the spectrum model. The values of $c_{0,1,2}$ are given for the wider ROI in KNM1,2 (\SIrange{14}{32}{\kilo\electronvolt}, entries \textit{knm12-1432-c0}, \textit{knm12-1432-c1}, \textit{knm12-1432-c2}) and for the narrower ROI in KNM3-SAP,3-NAP,4,5 (\SIrange{22}{34}{\kilo\electronvolt}, entries \textit{knm345-2234-c0}, \textit{knm345-2234-c1}, \textit{knm345-2234-c2}).

\noindent\textbf{Source potential.} The variance of the spectrum broadening due to the source-potential inhomogeneities is given by the entry \textit{SourcePotentialBroadeningSquared}. An effective energy-loss shift parameter $\Delta$ is presented by \textit{SourcePotentialEnergyLossShift}. Both parameters are given individually for KNM1,2,3-SAP and KNM3-NAP,4,5 due to different source densities. The $\Delta$ parameter is considered fully correlated between all campaigns. Finally, the long-term variation of the source potential is summarized by the entry \textit{PotentialDriftBroadeningSquared}, which provides the variance of the corresponding broadening. The variance is estimated independently for each campaign, as described in section~\ref{SubSec:Plasma}.